\documentclass[rsi,amsmath,amssymb,superscriptaddress,reprint]{revtex4-1}

\usepackage{graphicx}
\usepackage{dcolumn}
\usepackage{bm}
\usepackage[utf8]{inputenc}
\usepackage[T1]{fontenc}
\usepackage{mathptmx}

\usepackage{upgreek}
\usepackage[colorlinks=true,
linkcolor=MediumBlue,
urlcolor=MediumBlue,
citecolor=MediumBlue]{hyperref}

\usepackage[svgnames]{xcolor}

\begin{document}

\title{High sensitivity, levitated microsphere apparatus for short-distance force measurements}

\author{Akio Kawasaki}
\email{akiok@stanford.edu}
\affiliation{Department of Physics, Stanford University, Stanford, California 94305, USA}
\affiliation{W. W. Hansen Experimental Physics Laboratory, Stanford University, Stanford, California 94305, USA\looseness=-1}

\author{Alexander Fieguth}
\affiliation{Department of Physics, Stanford University, Stanford, California 94305, USA}

\author{Nadav Priel}
\affiliation{Department of Physics, Stanford University, Stanford, California 94305, USA}

\author{Charles P. Blakemore}
\affiliation{Department of Physics, Stanford University, Stanford, California 94305, USA}

\author{Denzal Martin}
\affiliation{Department of Physics, Stanford University, Stanford, California 94305, USA}

\author{Giorgio Gratta}
\affiliation{Department of Physics, Stanford University, Stanford, California 94305, USA}
\affiliation{W. W. Hansen Experimental Physics Laboratory, Stanford University, Stanford, California 94305, USA\looseness=-1}

\date{\today}

\begin{abstract}
A high sensitivity force sensor based on dielectric microspheres in vacuum, optically trapped by a single, upward-propagating laser beam, is described. Off-axis parabolic mirrors are used both to focus the 1064~nm trapping beam and to recollimate it to provide information on the horizontal position of the microsphere.  The vertical degree of freedom is readout by forming an interferometer between the light retroreflected by the microsphere and a reference beam, hence eliminating the need for auxiliary beams. The focus of the trapping beam has a 1/e$^2$ radius of 3.2~$\upmu$m and small non-Gaussian tails, suitable for bringing devices close to the trapped microsphere without disturbing the optical field.  Electrodes surrounding the trapping region provide excellent control of the electric field, which can be used to drive the translational degrees of freedom of a charged microsphere and the rotational degrees of freedom of a neutral microsphere, coupling to its electric dipole moment. With this control, the charge state can be determined with single electron precision, the mass of individual microspheres can be measured, and empirical calibrations of the force sensitivity can be made for each microsphere. A force noise of $<1\times10^{-17}$~N/$\sqrt{\rm Hz}$, which is comparable to previous reports, is measured on all three degrees of freedom for 4.7~$\upmu$m diameter, 84~pg silica microspheres.  Various devices have been brought within $1.6~\upmu$m of the surface of a trapped microsphere. Metrology in the trapping region is provided by two custom-designed microscopes providing views in the horizontal and one of the vertical planes.  The apparatus opens the way to performing high sensitivity three-dimensional force measurements at short distance.

\end{abstract}

\maketitle

\section{Introduction}
Force and acceleration measurements find numerous applications in science and technology.  In recent times, traditional sensors based on mechanical springs have been complemented by more elaborate systems based on non-contact electromagnetic effects.  These include atomic force microscopes~\cite{PhysRevLett.56.930}, optical~\cite{RepProgPhys.72.076901,ClassQuantumGrav.32.074001,ClassQuantumGrav.32.024001,NatAstro.3.2397} and atom interferometers~\cite{nature.400.849,Metrologia.38.25,PhysRevA.91.033629,PhysRevA.88.043610}, and falling corner-cube-gravimeters~\cite{Metrologia.32.159}, each emphasizing a different aspect of the measurement and generally achieving excellent sensitivity and accuracy.  Since the pioneering work of Ashkin~\cite{PhysRevLett.24.156,ApplPhysLett.19.283,ApplPhysLett.30.202,OptLett.11.288}, optical tweezers have also been used as force sensors, with initial applications to particles in liquid suspensions, primarily in biology~\cite{NatPhot.5.318, MethEnzymology.475.377, RevSciInstrum.75.2787} and polymer science~\cite{Science.264.819}.  In the last decade, the path to trapping dielectric microspheres (MSs) in high vacuum was established~\cite{NatPhys.7.527}, and various groups have used the technique to obtain highly sensitive force sensors that are well isolated from the environment~\cite{PhysRevA.93.053801,ApplPhysLett.111.133111,PhysRevA.99.023816,PhysRevLett.109.103603,IntJModPhys.B27.1330018,NatPhys.12.806,ApplPhysLett.111.133111,2001.1093,NatNanotechnol.2.89,Science.367.6480}.

Force sensors using optically trapped MSs have the ability to carry out measurements at distances of sub-millimeter scale.  This can be achieved by inducing a force between the MSs and specially designed sources of the desired interaction, placed in close proximity to the trapped MS. This has been demonstrated in a few cases~\cite{PhysRevA.99.023816,PhysRevLett.117.101101,PhysRevA.98.053831,PhysRevA.98.013852}, where separations between MSs and attractors have reached the scale of several micrometers.

In this paper, a multipurpose optical tweezer system, evolved from the apparatus  described in Ref.~\cite{PhysRevA.97.013842} and optimized to search for new fundamental interactions at the micrometer scale, is described. The system is currently used to trap silica MSs of diameters 4.7 and 7.6~$\upmu$m and has achieved separations of 1.6~$\upmu$m between the surfaces of a MS and a nearby device with nominal noise conditions.  The trap uses a single beam of 1064~nm wavelength with interferometric readout on all three degrees of freedom, as demonstrated in Ref.~\cite{PhysRevA.97.013842}.  A number of features, most notably a final focusing and recollimation employing off-axis parabolic mirrors, have been introduced to minimize beam halos at the focus, enabling closer access to the trapping region with minimal distortion of the optical field.  A large vacuum chamber allows for the introduction of several motorized actuators, important for the manipulation of devices near the trapped MSs under vacuum.  Three-dimensional metrology of these devices around the trap region is provided by two orthogonal microscopes.  The readout of the polarization state of the trapping light after its interaction with the MS is used to measure the rotation of trapped MSs, owing to their residual birefringence. Rotation of the MS can be induced by producing a rotating electric field that couples with the permanent electric dipole moment generally present in the silica MSs used~\cite{PhysRevA.99.041802}.  Trap stabilization against long term drifts of the interferometric platform affords a noise spectrum that is flat down to ${\sim}1$~Hz.  

Along with its primary motivation to search for new interactions at the micrometer scale~\cite{PhysRevLett.105.101101}, the system described may be used for the investigation of Casimir forces~\cite{Casimir,PhysRevLett.78.5,PhysRevLett.81.4549,PhysRevA.62.062104,PhysRevLett.81.4549,PhysRevLett.88.041804,PhysRevLett.91.050402,EPL.112.44001} and other applications~\cite{PhysRevLett.110.071105,ClassQuantumGrav.37.075002,PhysRevD.99.023005} requiring extreme force sensitivity.

\section{Optics Setup} \label{OpticsSetup}
\begin{figure}[!tb]
    \includegraphics[width=1\columnwidth, bb=0 0 802 210]{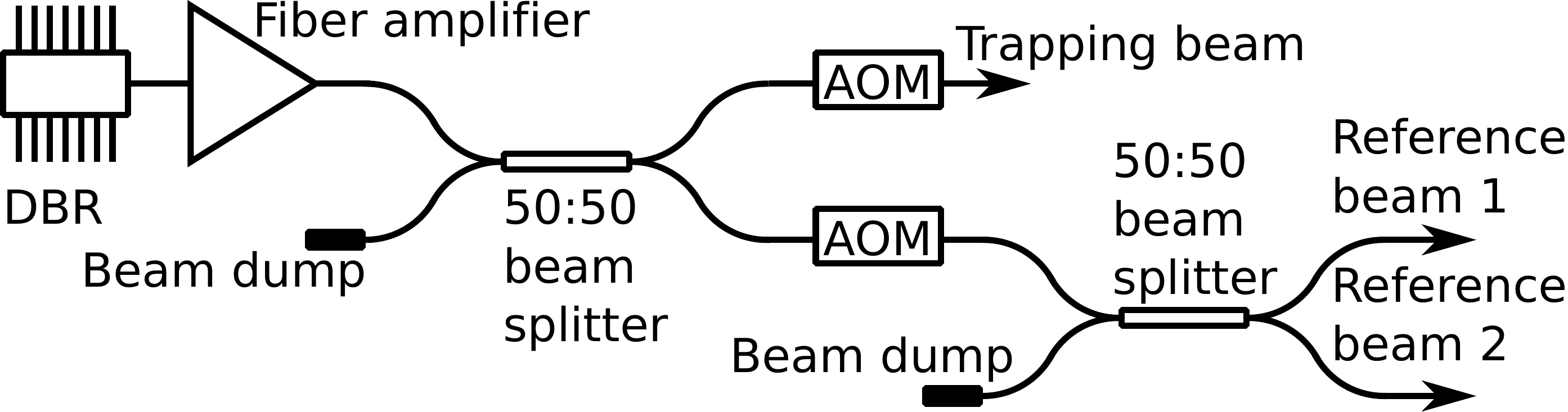}
    \caption{Schematic depiction of the laser system: all of the components shown are fiber-coupled with PM fibers, and arrows indicate the fiber outcouplers to free space, shown in Fig.~\ref{OpticsSystem}. DBR: distributed Bragg reflector laser and AOM: fiber-coupled acousto-optic modulator.}
    \label{LaserSystem}
\end{figure}

The 1064~nm trapping light is produced using a distributed Bragg reflector laser (Innolume LD-1064-DBR-150) to seed a ytterbium-doped fiber amplifier (Thorlabs YDFA100P), resulting in a maximum power of 100~mW. The production of the trapping beam and the reference beams for the heterodyne detection system makes use of fiber optic components based on single-mode PM980 fiber or equivalent, as shown in Fig.~\ref{LaserSystem}. Light from the fiber amplifier first goes through a 50:50 fiber-coupled polarization maintaining (PM) beam splitter. The two output channels of this splitter are independently frequency shifted by ${\sim}150$~MHz to a final frequency difference of 125~kHz, using two fiber-coupled acousto-optic modulators (AOMs, Gooch and Housego T-M150-0.4C2G-3-F2P). One channel, the trapping beam, is then launched to free space for further manipulation, while the second channel is subsequently split into two halves to produce reference beams for the detection of the vertical ($z$) and horizontal ($x-y$) MS positions. In order to passively stabilize its temperature, the entire fiber optics system is heat sunk to the 4,000~kg granite table on which the trap is located, and embedded in foam to decouple it from the ambient air, which help to reduce long term drift of the interferometric readout. Continuous temperature measurements show that the laboratory air conditioning system maintains the air temperature at $23.0~^{\circ}$C$\pm0.5~^{\circ}$C. 

\begin{figure*}[!tb]
    \includegraphics[width=1.9\columnwidth, bb=0 0 802 325]{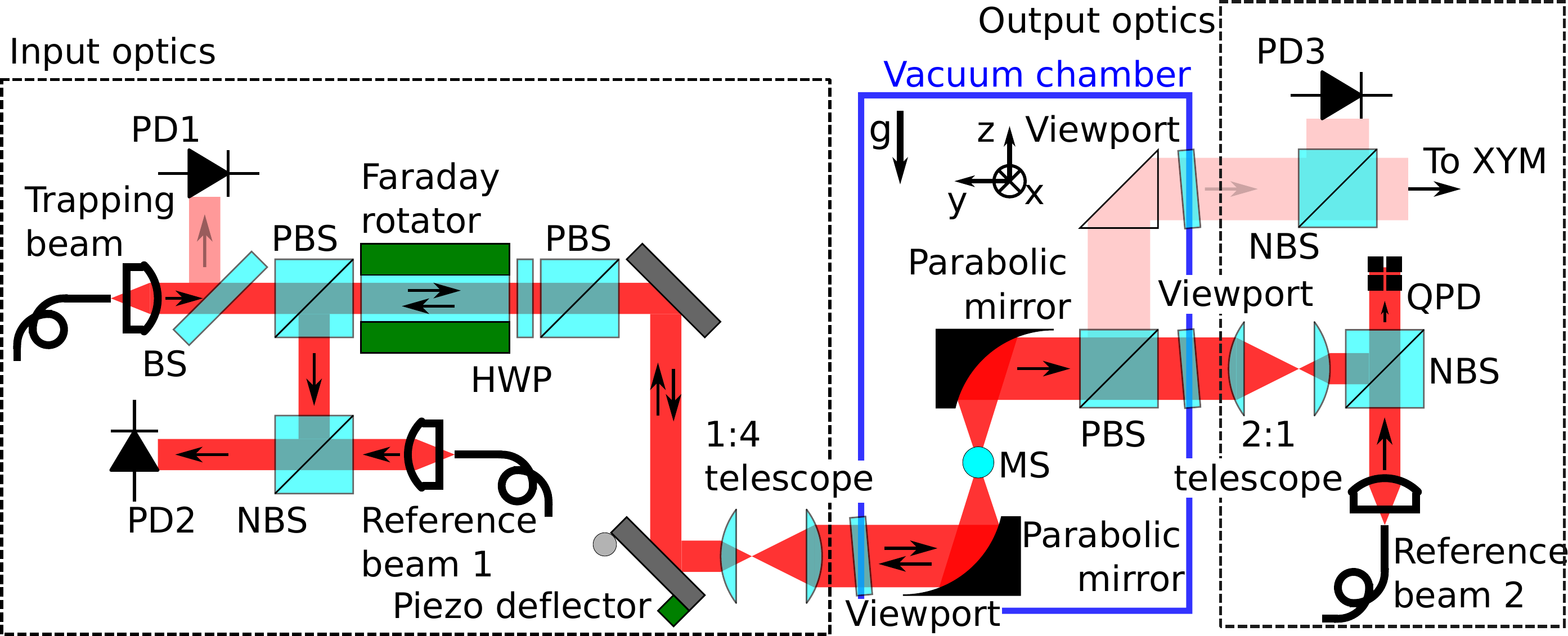}
    \caption{Simplified free-space optics system: only the essential components are drawn, and some mirrors are omitted. Each reference beam has a pair of mirrors (not shown) between the fiber outcouplers and the NBSs to align their wavefronts to that of the trapping beam. The parabolic mirrors and MS are shown from a side view, while the remainder of the components are shown from a top view. The two auxiliary microscopes described in Section~\ref{SectionCamera} operate independently from the trap and are not shown here. BS: beam sampler, PBS: polarizing beam splitter, NBS: nonpolarizing beam splitter (50:50), HWP: half waveplate, QPD: quadrant photodiode, PD1-3: photodiodes, and XYM: $x-y$ microscope.  
}
    \label{OpticsSystem}
\end{figure*}

Both the input and output free-space optical systems are each mounted on a $60\times30$~cm$^2$ breadboard. The breadboards themselves are the actuated elements of six-axis stages so that, once internal alignment between components on the breadboards is achieved, they can be collectively adjusted relative to the trap, which is directly mounted on the vacuum chamber on the granite table (see Section~\ref{Vacuum}). 

The trap and the surrounding optics are schematically illustrated in Fig.~\ref{OpticsSystem}.  The trapping beam is first launched to free space and collimated by a single aspheric lens (Thorlabs PAF2A-A15C), resulting in a 1/e$^2$ beam radius of $w=1.35$ mm. A part of the beam (${\sim}9\%$) is sent to a photodiode (PD, Thorlabs DET100A2) by a beam sampler (CVI W1-IF-1012-UV-1064-0) in order to monitor the power of the trapping beam. The rest of the beam passes through an optical isolator consisting of two polarizing beam splitters (PBSs, CVI PBS-1064-100), a Faraday rotator (Electro-Optics Technology 110-10299-0001-ROT), and a half waveplate (Newport 10RP02-34). The light is horizontally polarized after the optical isolator. 

The trapping beam is then sent to a piezoelectric deflector that steers the beam along two orthogonal axes and is used to provide feedback to the $x-y$ translational degrees of freedom. The deflector, placed in a Fourier plane of the trap, employs a 7~mm diameter mirror (Edmund Optics \#34-370) glued to a high bandwidth actuator (Thorlabs ASM003). The mechanical resonant frequency of this assembly is ${\sim}2$~kHz and feedback is applied predominantly at frequencies below ${\sim}1$~kHz. Finally, the beam is expanded with a 1:4 telescope using $d=5$~cm and $d=20$~cm lenses (Thorlabs AL2550H-B and Newport PAC32AR.16), with $d$ being the focal length, yielding $w=4.84$~mm, slightly smaller than the expected 5.4~mm. 

\begin{figure}[!tb]
    \includegraphics[width=1\columnwidth,bb=0 0 567 567]{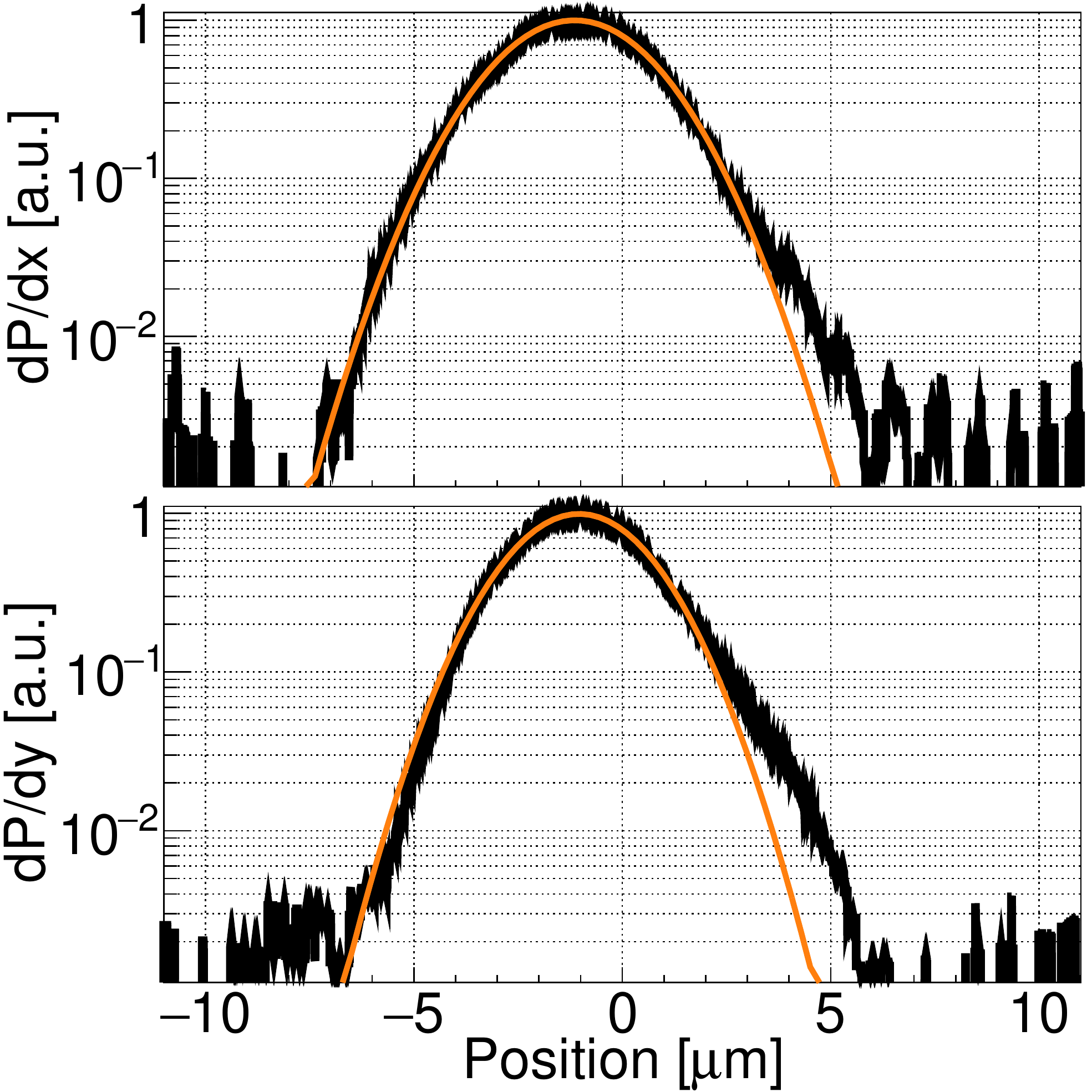}
    \caption{Beam profile at the trap for the $x$ and $y$ directions: the black lines are the data, and the orange curves are fits to Gaussian functions. The 1/e$^2$ beam radii are 3.4 $\upmu$m in the $x$ direction and 3.1 $\upmu$m in the $y$ direction. During the measurement, the nanofabricated device approaches the beam from the negative side. Coordinates of the $x$ and $y$ axes are determined by strain gauges in the piezoelectrically driven flexures (See Section \ref{NanoposStagePorts}).
 }
    \label{BeamProfileAtFocus}
\end{figure}

All vacuum chamber viewports are made from 5~cm diameter windows (CVI W2-IF-2037-UV-1064-0 for the trapping beam and CVI W2-IF-2037-UV-633-1064-0 for auxiliary imaging, see Section~\ref{SectionCamera}) custom-mounted with a tilt of 5$^{\circ}$ and sealed with O-rings. The trapping beam is focused inside the vacuum chamber by a $d=50.8$ mm off-axis parabolic mirror (Edmund Optics \#35-507). The mirror sits on a five-axis stage (Newport 9082-V), where rotation around the $x$ axis is the non-adjustable degree of freedom.   Reflective optics are preferred here, owing to the absence of spurious reflections from optical interfaces in refractive optics. The 1/e$^2$ beam radius at the trap focus $w_0$ is estimated with a knife-edge method, making use of a nanofabricated device mounted on the main nanopositioning stage (see Section~\ref{NanoposStagePorts}). The measured beam size is $w_0=3.4$~$\upmu$m (3.1 $\upmu$m) in the $x$ ($y$) direction, as shown in Fig.~\ref{BeamProfileAtFocus}, averaged to 3.2 $\upmu$m corresponding to a Rayleigh range of 31 $\upmu$m. This is in reasonable agreement with the theoretical prediction of 3.5~$\upmu$m from the measured numerical aperture of 0.095. A careful choice of the optical components, the use of 5~cm aperture optics after expanding the beam radius, and a careful alignment are important to achieve close to nominal performance. It is estimated that the residual imperfections and astigmatism ($x$ and $y$ foci displaced by ${\sim} 10~\upmu$m in $z$) are dominated by nonideal alignment.

The diverging beam emerging from the trapping region is recollimated by an identical parabolic mirror, also mounted on the same type of five-axis stage as the first parabolic mirror. The trapping beam exits the vacuum chamber horizontally polarized, while vertically polarized light is separated with a PBS (Edmund Optics \#65-606) and projected onto a PD (Thorlabs DET100A2) in order to monitor the MS rotation~\cite{PhysRevA.99.041802}. After being extracted from the vacuum chamber, the trapping beam passes through a 2:1 telescope composed of a $d=20$~cm and a $d=10$~cm lens (Newport PAC32AR.16 and Thorlabs LA1509-C), and is combined in a 50:50 non-polarizing beam splitter (NBS, Newport 10BC17MB.2) with a reference beam. The resulting superposition of light is projected onto a quadrant photodiode (QPD, Hamamatsu S5980). The 2:1 telescope coarsely matches the mode of light transmitted through the MS to that of the reference beam.  

The counter propagating beam, retroreflected by the surface of the MS, is extracted from the rejection port of the optical isolator on the input optics side, and combined with the second reference beam using a 50:50 NBS (Newport 10BC17MB.2) and projected onto a PD (Thorlabs DET100A2).  

Both reference beams are launched to free space by outcouplers with adjustable beam size and focal length (Thorlabs ZC618APC-C), which allow further optimization of the mode-matching to the transmitted and reflected components of the trapping beam. Prior to the NBSs, each reference beam undergoes two reflections so that the position and angle of incidence on the PD or the QPD can be adjusted. 

As mentioned, the output optics are also mounted on a breadboard that allows for the overall alignment with respect to the trap.  Both input and output free-space optics are enclosed in lens tubes to reduce air currents and microphonic effects (except for the two motorized mirrors in the reference beam path for the $x-y$ position detection). The entirety of each of the six-axis stages holding the breadboards is enclosed in boxes made of acrylic to further suppress noise.

\section{Data Acquisition and Feedback}
On the PD dedicated to the $z$ position of the MS (PD2 in Fig.~\ref{OpticsSystem}), and the QPD dedicated to the $x-y$ position of the MS, the incident optical power is modulated by the $\Delta f = 125$~kHz frequency shift between the trapping and reference beams.  The five photocurrents corresponding to PD2 and each quadrant of the QPD are individually amplified and then digitized at a sampling frequency $f_s = 500$~kS/s, exactly four times $\Delta f$. By phase-locking the radio frequency (RF) synthesizers driving the AOMs to the master clock driving the analog-to-digital converters (ADCs), real-time estimates of the amplitude and phase of the oscillating photocurrent are obtained. Every sample $A_i$ is spaced by a quarter wavelength, and thus,

\begin{align}
    A_i = G R_{\rm t} I_{\rm photo} \begin{cases} \text{sin}(\phi) & i=1,5,... \\ -\text{cos}(\phi) & i = 2,6,... \\ -\text{sin}(\phi) & i=3,7,... \\ \text{cos}(\phi) & i=4,8,... \end{cases}, \label{eq:demod}
\end{align}

\noindent where $G$ is the unitless voltage-gain of the amplifier circuit, $R_{\rm t}$ is the transimpedance resistance used to convert the photocurrent $I_{\rm photo}$ to a voltage, and $\phi$ is an arbitrary, but fixed, phase offset between the photocurrent and the digitizer's master clock. The phase and amplitude of this signal can be different for each quadrant and for the $z$-position PD. The amplitude $G R_{\rm t} I_{\rm photo}$ is estimated as $G R_{\rm t} I_{\rm photo} = (A_i^2 + A_{i-1}^2)^{1/2}$, while the phase is estimated as $\phi = \tan^{-1} \left[ A_1 / (-A_2) \right]$, or as $\phi = \tan^{-1} \left[ -A_3 / (-A_2) \right]$ etc., where the appropriate negation and ratio repeats every fourth sample, as seen in Eq.~(\ref{eq:demod}). This procedure is often referred to as digital demodulation.

Since displacements of the MS in the horizontal ($x-y$) plane at the trap produce displacements of the transmitted trapping beam in the plane of the QPD, while the reference beam is fixed, the $x-y$ degrees of freedom are read out as imbalances between the amplitude of interference photocurrent across the QPD, normalized by the total photocurrent amplitude. Vertical ($z$) displacements of the MS change the optical path length of the retroreflected light, which are read out directly from changes in the phase of the interference photocurrent from PD2.

The aforementioned demodulation and construction of the $x$, $y$, and $z$ signals takes place within a field programmable gate array (FPGA), embedded with ADCs (National Instruments PXIe-7858R). This FPGA also computes and generates the stabilizing feedback on similarly embedded digital-to-analog converters, which are sent to the two orthogonal axes of the piezoelectric deflector for $x$ and $y$ and to the amplitude modulation port of the RF synthesizer driving the trapping beam AOM for $z$. The FPGA is connected to a host computer to which it transfers the demodulated amplitudes and phases of PD2 and each QPD-quadrant, together with the position estimates and the generated feedback for offline analysis. While in the $x$ and $y$ directions interferometry is used only to suppress stray light from sources displaced from the trap center, it is an essential feature to measure the $z$ position that allows the trap to operate with a single beam, affording unimpeded access in the horizontal plane~\cite{PhysRevA.97.013842,PhysRevA.99.023816}.

Temperature drifts affect the optical path length of all light used for interferometry. In the $x$ and $y$ degrees of freedom, the amplitude of interference photocurrent in each quadrant of the QPD is, at first order, independent of the phase $\phi$. Fluctuations in the optical power are also suppressed by normalizing the $x$ and $y$ estimates by the total photocurrent amplitude. In the $z$ degree of freedom, optical path length fluctuations propagate directly into the estimate of the $z$ position. These fluctuations are attributed to both air currents and residual temperature drifts and have, roughly, a $1/f$ spectrum. They are suppressed passively by heat-sinking, the insulation, and the enclosures mentioned in the Sec. \ref{OpticsSetup}, and actively by using the image reconstructed by the $y-z$ microscope, described in Section~\ref{SectionCamera}.  This latter technique addresses temperature fluctuations with periods of minutes.

\section{Vacuum system and trap mechanics}\label{Vacuum}
The optical trap is located within a $42.5\times42.5\times34.26$~cm$^3$ cubical aluminum vacuum chamber sealed on all six sides with International Standard Organization (ISO) flanges and Viton gaskets, which simplify access to the trap. High vacuum is achieved primarily with a 250~l/s turbomolecular pump (TMP, Pfeiffer HiPace 300), roughed by a scroll pump (Edward XDS35i) located in a separate room to reduce the acoustic noise. Although the use of Viton gaskets limits the attainable vacuum level to ${\gtrsim} 10^{-8}$ hPa, the actual base pressure of $2.4\times10^{-7}$~hPa achieved after a few days of pumping is thought to be limited by the outgassing of the motorized stages, some of which have stated vacuum compatibility of $10^{-6}$ hPa. A system bakeout is impractical with the current setup. A high conductance port is available directly at the top of the chamber for the future installation of a large getter pump (SAES CAPACITORR HV1600) to improve the vacuum level. The current vacuum level does not limit the performance of the system, and data presented here are collected when the pressure reaches ${\lesssim} 10^{-6}$~hPa.

The bottom ISO-320 flange rests on an adapter to the granite table and contains a zero-length reducer from ISO-320 to ISO-100. The pumping system is connected to this flange and extends to the space below the granite table through a hole. A ceramic break (MDC 9632010) and a 10 cm long bellow are located between the pumping system and the main chamber for electrical and vibration isolation, respectively. The TMP can be isolated from the main chamber by a gate valve (MDC GV-4000M-P-01), while a custom tee between the chamber and gate valve connects a low conductance bypass for slow pumping at low vacuum with only the scroll pump. The bypass is throttled by a manual leak valve (Dunway VLVE-1000) and can be turned off by a pneumatically driven on-off valve (VAT 28324-GE41-0002/0068). Leaking N$_2$ gas into the chamber is accomplished with an electrically controlled leak valve (MKS 148JA53CR1M) together with a pneumatically driven on-off valve (US Solid PSV00032).

The vacuum system is monitored by a full range vacuum gauge (Pfeiffer PKR 251), as well as a residual gas analyzer (MKS eVision+). However, the use of these devices is found to affect the charge state of the trapped MSs, and therefore, during charge-sensitive measurements, a capacitance manometer (MKS 627FU2TLE1B), with a minimum measurable pressure of $2 \times 10^{-5}$~hPa, is used to monitor the vacuum level. All gauges are mounted on conflat (CF) flange ports on an adapter nipple at the top of the main chamber. The rotational dynamics of the MS can also be used to measure the vacuum level, as has been demonstrated with the predecessor apparatus~\cite{JVSTB.38.024201}.  

\begin{figure}[!tb]
    \includegraphics[width=1\columnwidth, bb=0 0 4032 3024]{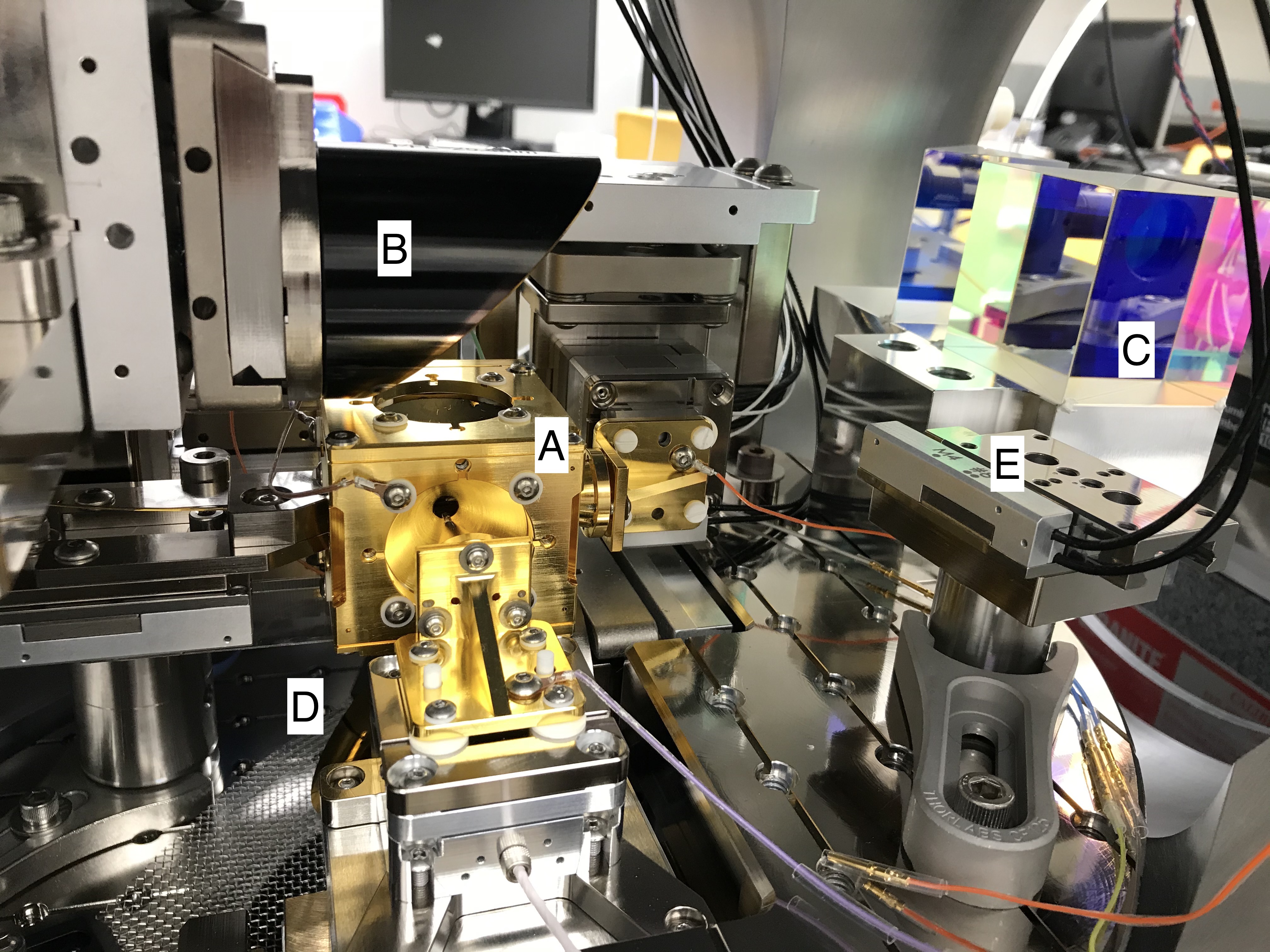}
    \caption{A photo of the components inside the vacuum chamber: the gold-coated cube in the middle (A) is the exterior of the six pyramidal electrodes surrounding the trapping region. The black diagonally-cut cylinder above the electrodes (B) is the recollimating parabolic mirror, which sends the collimated trapping beam to the PBS on the right (C). A small part of the optical surface of the focusing parabolic mirror is barely visible (D), under the cube. The motorized stage in the foreground of the PBS (E) supports the dropper (see Section~\ref{TrapOperation}), not installed in this image. }
    \label{InsidePhoto}
\end{figure}

Components inside the vacuum chamber are mounted on a 25.4 mm lattice of 1/4-20 screw holes on the interior of the bottom ISO-320 flange. A view of the components inside the chamber is shown in Fig.~\ref{InsidePhoto}. The trapping region is surrounded by six identical electrodes shaped as truncated pyramids.  The electrode faces are 4.3~mm away from the trap center, forming a cubical cavity of 8.6~mm side, with narrow gaps between electrodes.  Each electrode is hollowed out and, on the trap end, terminates with a 5.3~mm diameter aperture providing optical and mechanical access to the center. A cross section in the $x-y$ plane and at the nominal $z$ position of the trap is shown in Fig.~\ref{TrapRegionFig}. In addition to the shielding against stray electric fields, each electrode is electrically isolated and can be independently biased in order to apply electrical forces and torques to the MSs. Torque can be applied with a constant magnitude rotating electric field, generated by four phased sinusoids on four coplanar electrodes, that couples to the residual electric dipole moment generally present in the silica MSs used here~\cite{PhysRevA.99.041802,JVSTB.38.024201}. The electric field within the trapping region is calculated using finite element analysis for any configuration of electrode biases. The entire electrode assembly and mounting structure are composed of 6061 aluminum alloy and gold coated by electrode-poised plating. Surfaces facing the trap region, as well as the conical cavities of the top and bottom electrodes through which the trapping beam propagates, are further coated with colloidal graphite (Electron Microscopy Sciences 12660) to reduce the scattering of stray light.

\section{Nanopositioning Stage Ports} \label{NanoposStagePorts}
\begin{figure}[!tb]
    \includegraphics[width=1\columnwidth,bb=0 0 604 487]{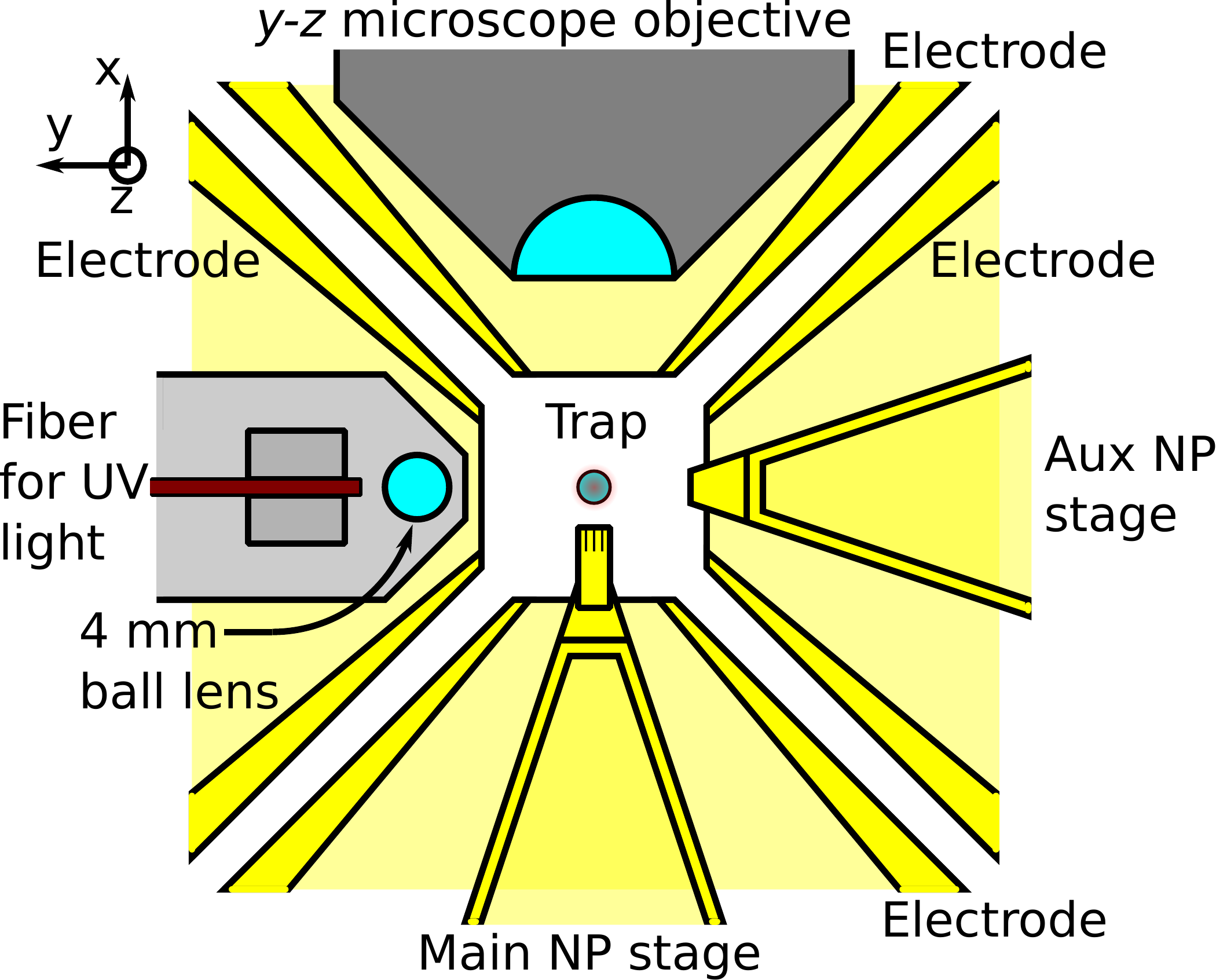}
    \caption{Components surrounding the trap in the $x-y$ plane: the main NP stage carries the nanofabricated device described in Section~\ref{NanoposStagePorts} mounted on its end. The y-z microscope objective is described in Section~\ref{SectionCamera}, and the fiber for UV light is described in Section~\ref{TrapOperation}. (Aux) NP stage: (auxiliary) nanopositioning stage.}
    \label{TrapRegionFig}
\end{figure}

The system is designed to provide stable and reproducible access to the trapping region for devices with dimensions in the $\upmu$m to mm range. Mechanical access is realized through the holes in two of the four electrodes in the horizontal plane, shown at the bottom and right of Fig.~\ref{TrapRegionFig}. In the current configuration, these house the main nanopositioning stage and an auxiliary nanopositioning stage, respectively. 

\begin{figure}[!tb]
    \includegraphics[width=1\columnwidth, bb= 0 0 1615 1104]{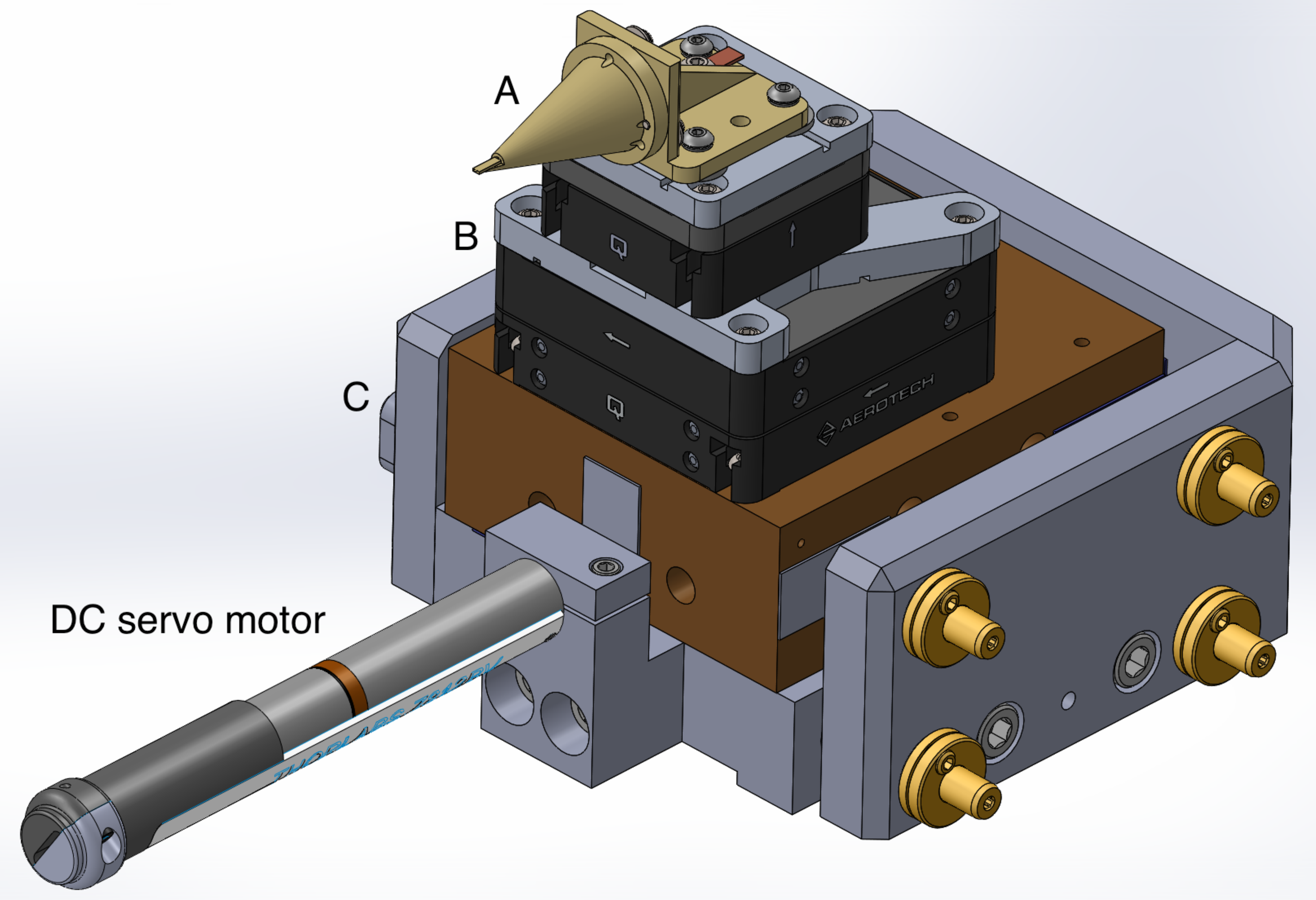}
    \caption{CAD model of the conical cantilever (A) onto which the primary silicon device is mounted, the piezoelectrically driven flexures (B), and the custom six-axis stage (C) for coarse alignment and device insertion.}
    \label{AttractorMount}
\end{figure}

The main nanopositioning stage consists of a stack of piezoelectrically driven flexures (Aerotech QNP40Z-100 for $z$ mounted on top of Aerotech QNP60XY-500 for $x-y$), which has a full range of $500~\upmu$m in $x$ and $y$, and $100~\upmu$m in $z$, with a resolution better than 1~nm and reproducibility of ${\sim}2$~nm. The actual positions along each axis are measured by strain gauges within the actuator and subsequently recorded together with the MS position. This actuator is mounted on a custom six-axis stage for coarse alignment to the trap center, with a long travel along the $x$ axis, required for insertion of the primary device into the trapping region. Four degrees of freedom are manually operated, and one degree of freedom is actuated with a piezoelectric motor (Newport 8301-UHV), while the translation along the $x$ axis is accomplished with a DC servo motor with 12~mm range (Thorlabs Z812V) for high repeatability. On top of the stack of piezoelectrically driven flexures, a gold-coated aluminum conical cantilever is mounted. To reduce mechanical load and optimize the bandwidth of the flexure's motion, the conical cantilever is hollowed out with a final wall thickness of 0.66 mm resulting in a mass of 3.13 g. The entire assembly is shown in Fig.~\ref{AttractorMount}. With the cantilever installed, the piezoelectrically driven flexure has a measured bandwidth of ${\sim}80~(100)$ Hz in the $x$ ($y$) direction, limited by mechanical resonances, and an estimated bandwidth of ${\sim}500$~Hz in the $z$ direction.

The auxiliary nanopositioning stage consists of a three-axis piezoelectrically driven flexure (Newport NPXYZ100SGV6) with $100~\upmu$m range in each of the three orthogonal directions.  For insertion into the electrode structure, a motorized stage with 12~mm range (Newport AG-LS25V6) is mounted on top of the three-axis actuator. A second, identical conical cantilever is mounted on this motorized stage for a secondary device to be inserted into the the trap orthogonally to the main device. Further coarse alignment and angular adjustment are provided by another manually adjusted platform (Newport 9071-V) onto which this assembly is mounted. The secondary system is intended for static use.

Nanofabricated devices, e.g. Ref.~\cite{AttractorPaper} are mounted at the ends of the two cantilevers with a conductive epoxy (Epo-Tek H21D). Typical devices have a high aspect ratio, being wide and long in the two horizontal dimensions (${\sim}1~$mm) and thin (10$~\upmu$m to 25$~\upmu$m) in the vertical dimension, in order to minimize their effect on the converging and diverging trapping beam. Devices are typically gold-coated to optimally define their electrical potentials (although charge patches are still present~\cite{PhysRevA.99.023816}). Both devices can be independently biased, allowing the production of electric fields with large gradients in the immediate vicinity of a trapped MS~\cite{PhysRevA.99.023816,PhysRevLett.117.101101}. Each can also be used as a knife edge to scan across the beam and characterize its radius, center position, and focal point, as noted in Section~\ref{OpticsSetup} (see Fig.~\ref{BeamProfileAtFocus}) and demonstrated with a previous system~\cite{PhysRevA.97.013842}. Typically with this technique, the center position of the trapping beam relative to the nanofabricated device is determined with a precision of ${\sim} 0.1 ~ \upmu$m, where the device's position is measured with strain gauges within the piezoelectrically driven flexure on which it is mounted. The coordinate system of the device is then registered to the position at which the edge of the device crosses the beam focus (See Fig. \ref{BeamProfileAtFocus}). Additionally, the size of these particular MSs is known with a precision of 0.04~$\upmu$m [49]. Thus, the distance between the surface of the sphere and the surface of the nanofabricated device is determined by the relative positions of the trap and the nearby device, and the size of the MS, assuming the MS is trapped at the beam focus. The closest separation achieved between the surface of a MS and a nearby device is 1.6 $\upmu$m, at which position the device can be translated in front of a stably trapped MS. The orientation and position of the devices on the two nanopositioning stages relative to the trap are determined by a combination of their interactions with the trapping beam and the auxiliary $y-z$ and $x-y$ microscopes, as discussed in Sec. \ref{SectionCamera}. The relative position between the two devices can be registered by bringing them into contact, which is clearly visible by their elastic deformation. This has been found to be reproducible to within ${\lesssim}1~\upmu$m.

\section{Auxiliary Imaging Systems} \label{SectionCamera}
\begin{figure}[!tb]
    \includegraphics[width=\columnwidth, bb=0 0 960 769]{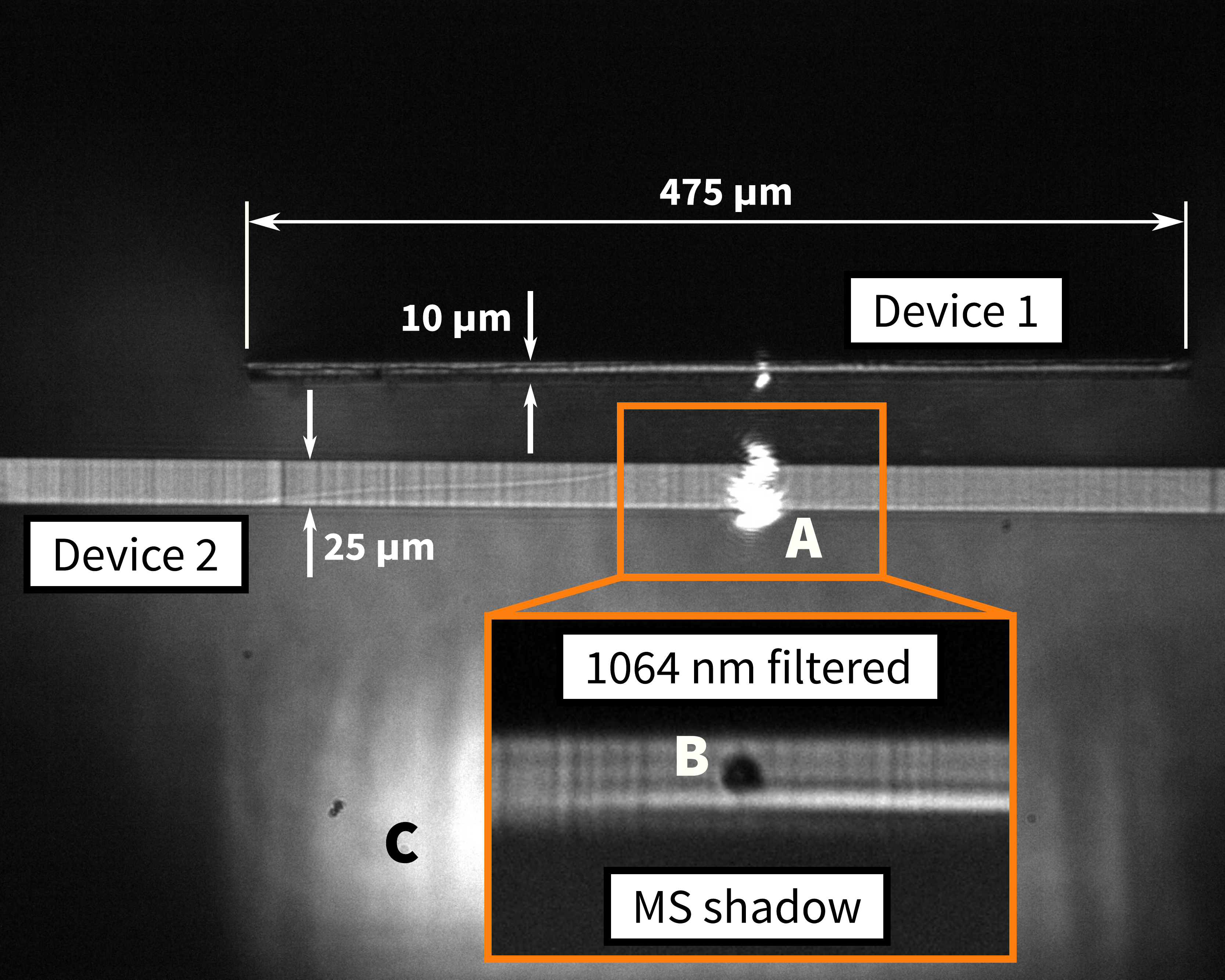}
    \caption{Composite image of two nanofabricated devices in the vicinity of the trap, together with a trapped MS, captured by the $y-z$ microscope. Device~1 is a silicon and gold beam~\cite{AttractorPaper}, mounted on the main nanopositioning stage. Device~2, mounted on the auxiliary nanopositioning stage, is a $1000~\upmu$m long all-silicon device, which has an L-shaped cross-section when viewed in the $x-z$ plane but not visible here, which usually houses device~1, shielding the trapped MS from background forces associated with device~1. In this image, device 1 is translated vertically from its nominal position and is out of focus, the latter reducing the apparent vertical extent. The main frame of the image shows the scattering of the trapping laser by the MS (A), which leads to saturation of several pixels of the camera. The inset is taken with a notch filter that has an optical depth of 6 for 1064~nm light (Thorlabs NF1064-44), in order to demonstrate the shadow of the MS (B) blocking the 870~nm illumination of device~2. The bright and blurry region (C) behind the device is caused by a reflection of the illumination from another part of device 1.}
    \label{SideMicroscopeImage}
\end{figure}

The setup is equipped with two separate systems for auxiliary imaging and metrology near the trapping region. The primary system images the $y-z$ plane of the trap through one of the electrodes in the $x-y$ plane, as shown in Fig.~\ref{TrapRegionFig}.  An infinity-corrected microscope objective (Nikon N10X-PF) is mounted, in vacuum, on a motorized linear stage with 12~mm range (Newport AG-LS25V6) to adjust the focal plane.  The image from the objective is extracted through a viewport and focused by a $d=20$~cm lens onto a CMOS camera (Allied Vision MAKO U-030B). A trapped MS is visible in Fig.~\ref{SideMicroscopeImage} due to scattered light from the 1064~nm trapping beam. In the same image, the nanofabricated devices near the trap are illuminated with an 870~nm wavelength light-emitting diode (LED) (Thorlabs LED870E), injected into the imaging system through a 50:50 NBS (Newport 10BC17MB.2) located in the path between the microscope objective within the vacuum chamber and the $d=20$~cm lens. Additionally, filters can be inserted in the optical path to selectively attenuate different wavelengths and optimize the visibility of various components. The combination of the illumination and a filter of 1064 nm light provides an additional way of imaging the MS, which casts a shadow on the device surfaces behind it.

While one pixel of the camera spans $0.5~\upmu$m in the image plane, the resolution of the imaging system is close to the diffraction limit ($1.5~\upmu$m for 870~nm light), verified using a USAF1951 resolution test chart. The entire imaging system can be cross-calibrated into the coordinate system of the piezoelectrically driven flexures by imaging objects of known size (e.g. the nanofabricated devices) within the trapping region. 

This $y-z$ microscope is also used in a slow loop of the feedback maintaining a constant vertical position of the MS. The stability of this slow feedback is better than $1~\upmu$m with a single image with an exposure time of 44 $\upmu$s taken every 10 s. 

The second imaging system provides a view of the $x-y$ plane at the $z$ position of the trap through the path for vertically polarized light, as shown in Fig.~\ref{OpticsSystem}. Half of the vertically polarized light from the output parabolic mirror is focused through a $d=7.5$~cm lens into an infinity-corrected objective, mounted on a manual linear translational stage to adjust the focal plane. The system is designed such that the objective can be switched to any model with a diameter of less than 3~cm, though a $\times 10$ magnification objective (Reichert Neoplan 1754) with a tube length of 16~cm is currently in use. Therefore, a broad range of magnifications are achievable. The image from the objective is focused by a second $d = 3$~cm lens onto another CMOS camera (Allied Vision MAKO U-029B).  The field of view is illuminated by 870~nm light produced by an LED (Thorlabs LED870E) and injected in the system through a NBS. The resolution in the center of the image is $\sim1.5~\upmu$m, near the diffraction limit. However, substantial aberrations are present away from the center of the image, owing to the lack of correction for the parabolic element and possible imperfections in the system alignment. For this reason, the $x-y$ imaging is used only for qualitative assessment and rough alignment. An improved microscope, using custom correction optics, may be installed at a later time.

\section{Trap Operation}\label{TrapOperation}
Dielectric MSs are prepared by rubbing a fused silica cover slip (dropper) on a powder of MSs laying on a sheet of glass. The dropper is glued to a piezoelectric transducer (Thorlabs PA4DG) and inserted between the recollimating parabolic mirror and the electrodes, with the face coated with MSs oriented down. The dropper is then vibrated by driving the piezoelectric transducer with an oscillating voltage, chirped between 150~kHz and 400~kHz. The MSs, held on the surface of the dropper by van der Waals forces, are released by the vibration, which is expected to generate kilo-$g$ or larger accelerations \cite{TLiThesis}. One dropper can be used to refill the trap many times, provided that the drive amplitude is gradually increased, suggesting that MSs are bound to the dropper with varying force. Between the initial preparation of the dropper and its depletion, the RF power driving the transducer typically has to increase from 1 mW to 1 W.

\begin{figure}[!tb]
    \includegraphics[width=1\columnwidth]{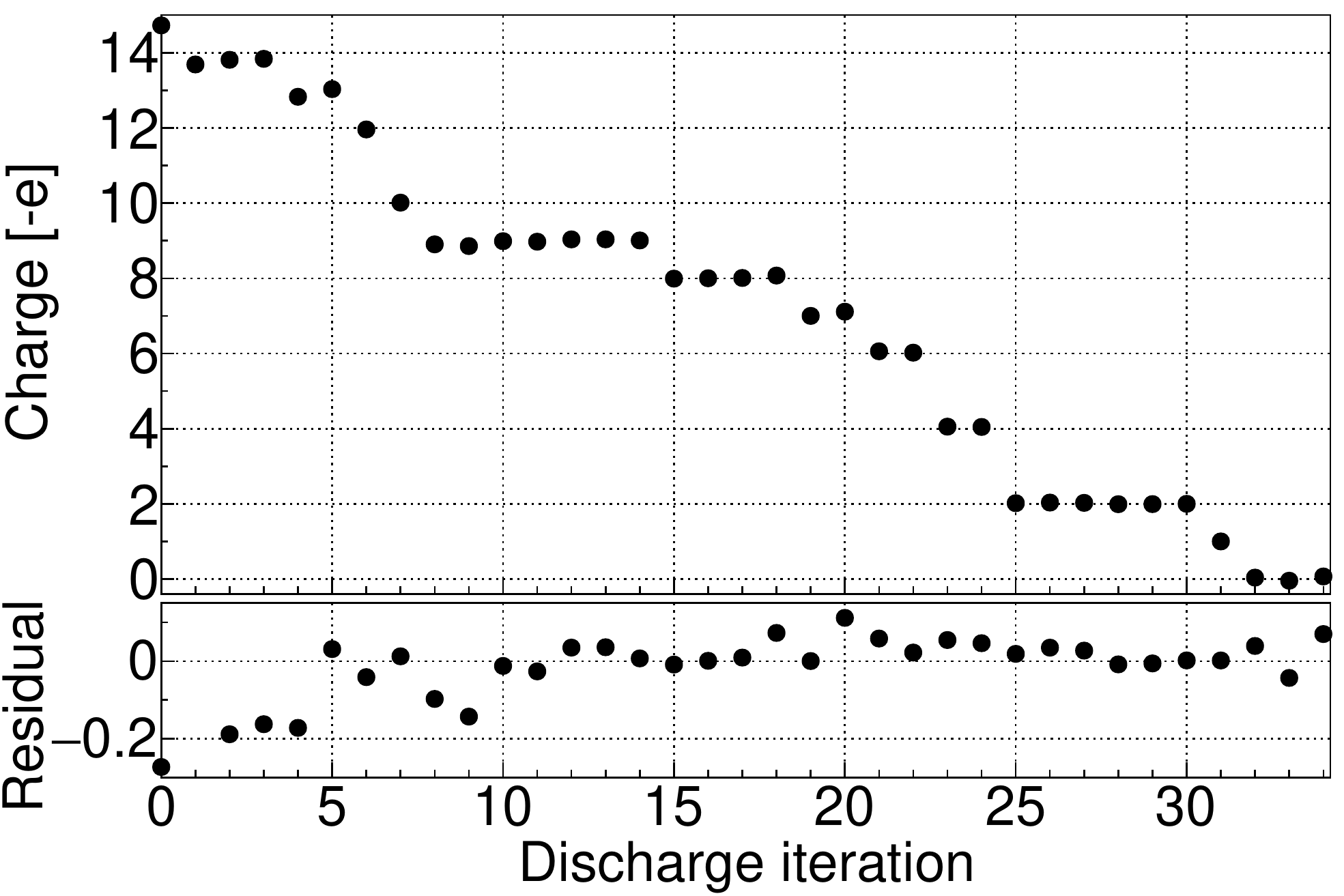}
    \caption{Typical discharging process: each data point corresponds to a 10-second measurement of the charge state (i.e. the amplitude of a MS response to a driving field), between which a certain number of flashes of the UV lamp occur. Quantization of the charge state in units of $e$ is observed.  When near the neutral charge state, the resolution of the measurement can be increased by increasing the driving voltage.}
    \label{Discharge}
\end{figure}

The loading process has a small but finite efficiency as most MSs fall without being captured in the trap.  Typically, ${\sim}6$~hPa pressure of N$_2$ is used to slow the falling MSs and make trapping possible, together with an increased laser power to increase the depth of the optical trap. For example, $4.7~\upmu$m diameter silica MSs~\cite{bangs_laboratories}, with mass ${\simeq}84$~pg~\cite{PhysRevApplied.12.024037} are caught with ${\sim}9$~mW of power in the trapping beam, while they are maintained at the focal point of the trap with ${\sim}1.7$~mW. Additionally, 7.6~$\upmu$m diameter, 420~pg mass silica MSs~\cite{German_sphere} are stably trapped with ${\sim}15$~mW of power and ${\sim}18$~hPa of N$_2$ pressure. It is expected that even larger MSs could be trapped with increased laser power. In order to minimize accumulation of MSs on the focusing parabolic mirror, a second, larger silica plate (catcher) is inserted between the electrode and the focusing parabolic mirror during loading.  The dropper and catcher are inserted and removed independently using motorized stages (Newport AG-LS25-27V6).  The trapped MS remains stable in the trap during the removal of the catcher, which is positioned 27 mm below the trap's focus, without any special effort. 

Once a MS is trapped, the low conductance bypass system is used to slowly pump down the vacuum chamber to 0.5~hPa (typically over 25~min), the pressure at which the feedback is initialized, typically with a reduced power of 1.8~mW (for $4.7~\upmu$m diameter MSs). Feedback from the position measurements of the three degrees of freedom of the MS is then applied to the laser power ($z$) and the pieozelectric deflector ($x-y$). The feedback for the vertical ($z$) degree of freedom includes proportional, integral, and derivative terms, and is applied in addition to an independent laser power stabilization which is primarily an integral term together with a proportional term. 
The slow, $z$ feedback mentioned in Section~\ref{SectionCamera} is only applied during long term measurements that typically last hours or days. The feedback for the horizontal ($x-y$) degrees of freedom is applied after the $z$ direction is stable and includes only derivative terms in order to damp the MS's resonant motion in the horizontal plane.

\begin{figure*}[!tb]
\begin{center}
    \includegraphics[width=2\columnwidth, bb=0 0 1440 540]{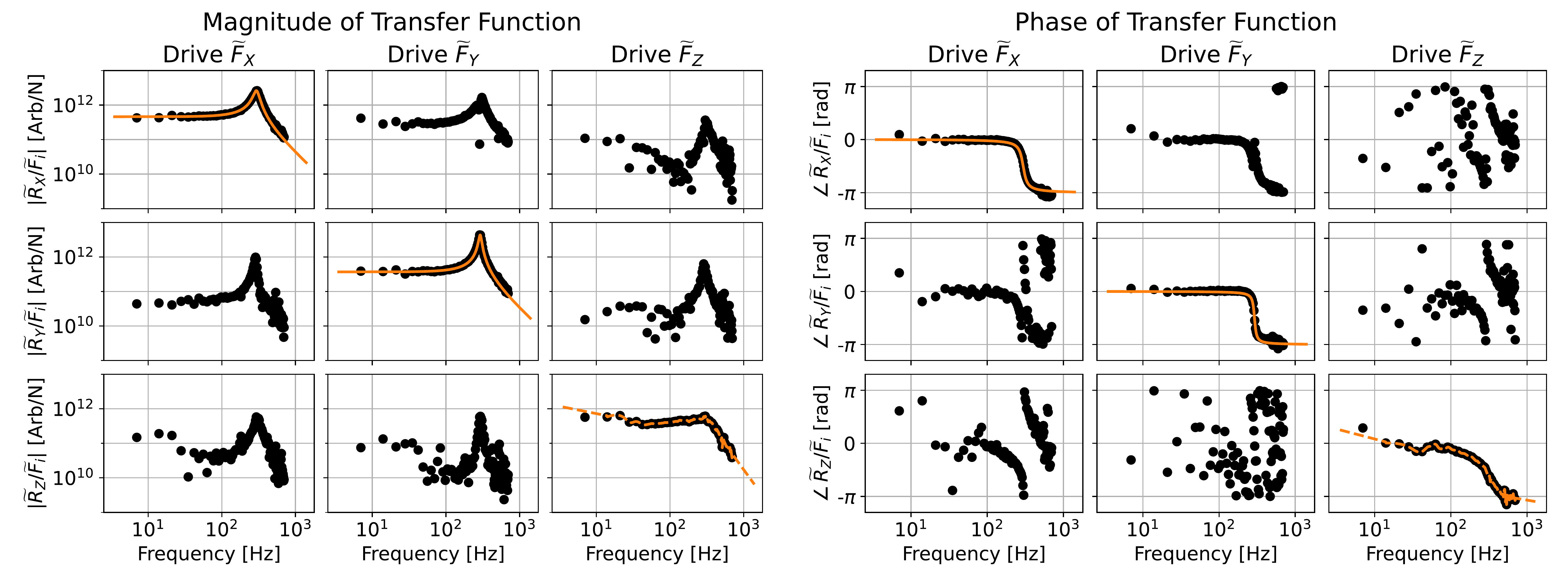}
\caption{Transfer functions of a typical MS. Shown are the magnitude (left) and phase (right) of the complex, frequency-dependent matrix $\mathbf{H}_{ij}(f)$ discussed in Section~\ref{ForceCalibration}. The orange solid curves for $(i,j)=(x,x), (y,y)$ are fits of the measured response to that of a damped harmonic oscillator, whereas the dashed line for $(i,j) = (z,z)$ is a quadratic spline interpolation. Off-diagonal interpolations are not shown. Resonant frequencies $f_{ii}$ and damping coefficients $\Gamma_{ii}$ obtained from the fit are $f_{xx}=301$~Hz$\pm1 $~Hz, $\Gamma_{xx}= 50$~Hz$ \pm 2$ Hz, $f_{yy}=292$~Hz$ \pm1 $~Hz, and $\Gamma_{yy}= 27$~Hz$ \pm 2$~Hz.}    \label{TransFunc}
\end{center}
\end{figure*}

The $z$ direction stabilization also allows the $z$ position of the MS to be varied with respect to the trap focus and other mechanical devices in its proximity. At the focus, where the trapping beam intensity is greatest, the optical spring constant confining the MS in the $x$ and the $y$ degrees of freedom has a maximum. The stochastic force on the MS from the 0.5~hPa of residual gas impacting the MS has a white frequency spectrum that drives these degrees of freedom, resulting in MS motion with a frequency spectrum well modeled with a driven, damped harmonic oscillator with a clearly observable resonant frequency. For a given $z$ position, the slight astigmatism produces two distinct values of $f_x$ ($f_y$), the resonant frequencies for the $x$ ($y$) directions, and the final setpoint is determined empirically to minimize the difference between $f_x$ and $f_y$. For the $4.7~\upmu$m diameter silica MSs, typical values are $f_x \simeq f_y \simeq 300$~Hz and $|f_x - f_y| \simeq 10$~Hz. At this location, the harmonic trap for $z$ is generated by the feedback, with $f_z\simeq 300$ Hz.

With $x$, $y$, and $z$ feedbacks on, the vacuum chamber is further pumped down, first with the low conductance bypass and then by opening the gate valve and starting the TMP at $<0.1$ hPa.  The system reaches a pressure of ${\sim}10^{-6}$~hPa in a few minutes, at which point the noise floor of the MS position is no longer dominated by the Brownian motion of the MS. MSs trapped in this fashion have been observed to be stable in the trap indefinitely in the absence of external disturbances.

The charge state of the trapped MS is measured by applying a sinusoidally oscillating voltage difference to a pair of opposing electrodes, and measuring the amplitude of the MS response to this driving field. For sufficiently long integration times, charge quantization is observed with a signal to noise ratio of ${\sim}20$. By increasing the amplitude of the driving field, the neutral state can be characterized exceedingly well as demonstrated in a simpler version of the trap in Ref.~\cite{PhysRevLett.113.251801}.  

As they are loaded in the trap, MSs can have either positive or negative overall charge. The charge state of MSs can be changed by flashing ultraviolet (UV) light from a xenon flash lamp (Hamamatsu L9455-13) into the trapping region. The light is brought into the trap with a multimode solarization-resistant UV fiber (Thorlabs UM22-600) and coupled to free space with a 4~mm fused silica ball lens (Edmund Optics \#67-385), as shown in the left of Fig.~\ref{TrapRegionFig}. The fiber is brought into the vacuum chamber using the feedthrough discussed in Ref.~\cite{ApplOpt.37.1762}. This system is capable of both increasing and decreasing the MS charge state.  Pulsing the UV lamp with nothing in close proximity to a MS tends to eject electrons from the MS, yielding a more positive charge state. If the sputtered gold surface of a nanofabricated device is placed behind the MS, more electrons appear to be ejected from the gold surface, changing the charge of the MS in the opposite direction. MS charge states upwards of $\pm 500~e$, with $e$ the fundamental charge can be obtained in this way or, if desired, net neutrality can be achieved. A typical discharging cycle is shown in Fig.~\ref{Discharge}. Since conductive structures close to the MS distort the electric field, absolute charge calibration is only performed as the charge increases by removing electrons from the MS, when the nanofabricated device is fully retracted.

\section{Force Calibration} \label{ForceCalibration}
With charge quantization, it is possible to empirically calibrate the response of the MS without assumptions. This is achieved by applying a sinusoidally oscillating electric field $E=E_0 \sin (2\pi f t)$ oriented along the degree of freedom to be calibrated, with $E_0$ being the amplitude of the electric field and $f$ being the frequency of the oscillation. The electric field amplitude $E_0$ at the MS location is calculated by finite element analysis, starting from the electrode geometry and the applied voltage. The force applied is then $F=q_{MS} E$, where $q_{MS} = ke$ with $e$ being the fundamental charge is deduced by counting the number of quanta $k$.  Simultaneously, the response of the MS, $R$, as determined by the imaging and demodulation procedure described previously, is also known. Thus, for a given frequency of an oscillating electric field, the ratio $R/F$ between the amplitude of MS response in arbitrary units and the amplitude of the applied force in physical units is derived. The same procedure also extracts any phase shift of the response relative to the drive. This can be done independently for each degree of freedom $x$, $y$, and $z$.

To measure the frequency dependence of this calibration, the electric field is applied in the form of a frequency comb 
\begin{equation} \label{FreqComb}
E(t) = E_0 \sum^{N}_{n=1} \sin  \left(2\pi n f t + \phi_n \right)
\end{equation}
\noindent with $\phi_n = 2\pi n^2/N$ being a phase shift, implemented in order to avoid large spikes of the electric field (consider the Fourier series of a delta function). Typically, $f=7~$Hz, $N=100$, and $E_0 = 100$~V/m are used for the initial characterization of the system. This electric field drive and the corresponding MS response can be continued indefinitely and averaged until a desired precision is achieved.

This method also provides a means to measure any cross talk inherent to the imaging, as well as the trap itself. With the symmetric electrode geometry employed here, forces are induced along specific coordinate axes, while the MS response along all three degrees of freedom is measured. Writing the applied force as $F_j = q_{\rm MS} E_j$ where $j \in \{x, y, z\}$, and $E_j$ of the form of Eq.~(\ref{FreqComb}), and writing the MS response as $R_i$ for $i \in \{ x, y, z \}$, each has a Fourier transform $\widetilde{F}_j(f)$ and $\widetilde{R}_i(f)$, respectively. The frequency dependent transfer function $\mathbf{H}_{ij}(f)$ is calculated as $\mathbf{H}_{ij} (f) = \widetilde{R}_i (f) / \widetilde{F}_j (f)$. The complex-valued $\mathbf{H}_{ij}(f)$ naturally includes amplitude ratios as well as phase shifts for each drive frequency. 

For a given $ij$, the amplitude and phase of $\mathbf{H}_{ij}(f)$ are smoothly interpolated (and extrapolated under certain assumptions) for frequencies not explicitly included in Eq.~(\ref{FreqComb}). Then, for any given measurement, the force applied to the MS is first constructed in the frequency domain from the measured MS response as $\widetilde{F}_j (f) = \mathbf{H}^{-1}_{ij} (f) \widetilde{R}_i (f)$. With sufficient optical alignment, the off-diagonal components of $\mathbf{H}_{ij}$ are typically five times smaller than their diagonal components. An estimate of the physical displacement of the MS at low frequencies is then obtained as $\widetilde{x}_j (f) = (1 / m_{\rm MS} \omega_j^2) \widetilde{F}_j(f)$, with $\omega_j$ being the angular resonant frequency for a particular coordinate axis, expressed in rad/s.

An example of $\mathbf{H}_{ij}$ is shown in Fig.~\ref{TransFunc}. In both the $x$ and the $y$ (horizontal) directions, the optical spring constant $k_j = m_{\rm MS} \omega_j^2$ is $\sim3 \times 10^{-7}~$N/m, while the translational damping constant $\Gamma$ is approximately $\sim50$~Hz and can be adjusted over a wide range by the active feedback. The resonant frequency and optical spring constant are in agreement with estimates from the Optical Tweezers Computational Toolbox~\cite{ott}, which assumes perfectly spherical MSs with uniform density, as well as an ideal Gaussian beam and diffraction-limited focus. In the $z$ (vertical) direction, relatively large proportional and integral gain terms are required to stabilize the MS's position, and a traditional damped harmonic oscillator model is insufficient to describe the frequency spectrum of the MS motion.

Between successive 10-s integrations, the response of the MS is stable to within 5\%, and reproducible on the timescale of days. The response of a charged MS to an oscillating electric field is linear up to $\sim2 \times 10^{-13}~$N of applied force, over a wide range of charge states $|q_{\rm MS}| \leq 500~e$. The linearity of the system described here matches or exceeds that of the previous version of the apparatus \cite{PhysRevA.99.023816}.

\section{Sensitivity} \label{SensitivitySection}
\begin{figure}[!tb]
     \includegraphics[width=1\columnwidth, bb=0 0 567 567]{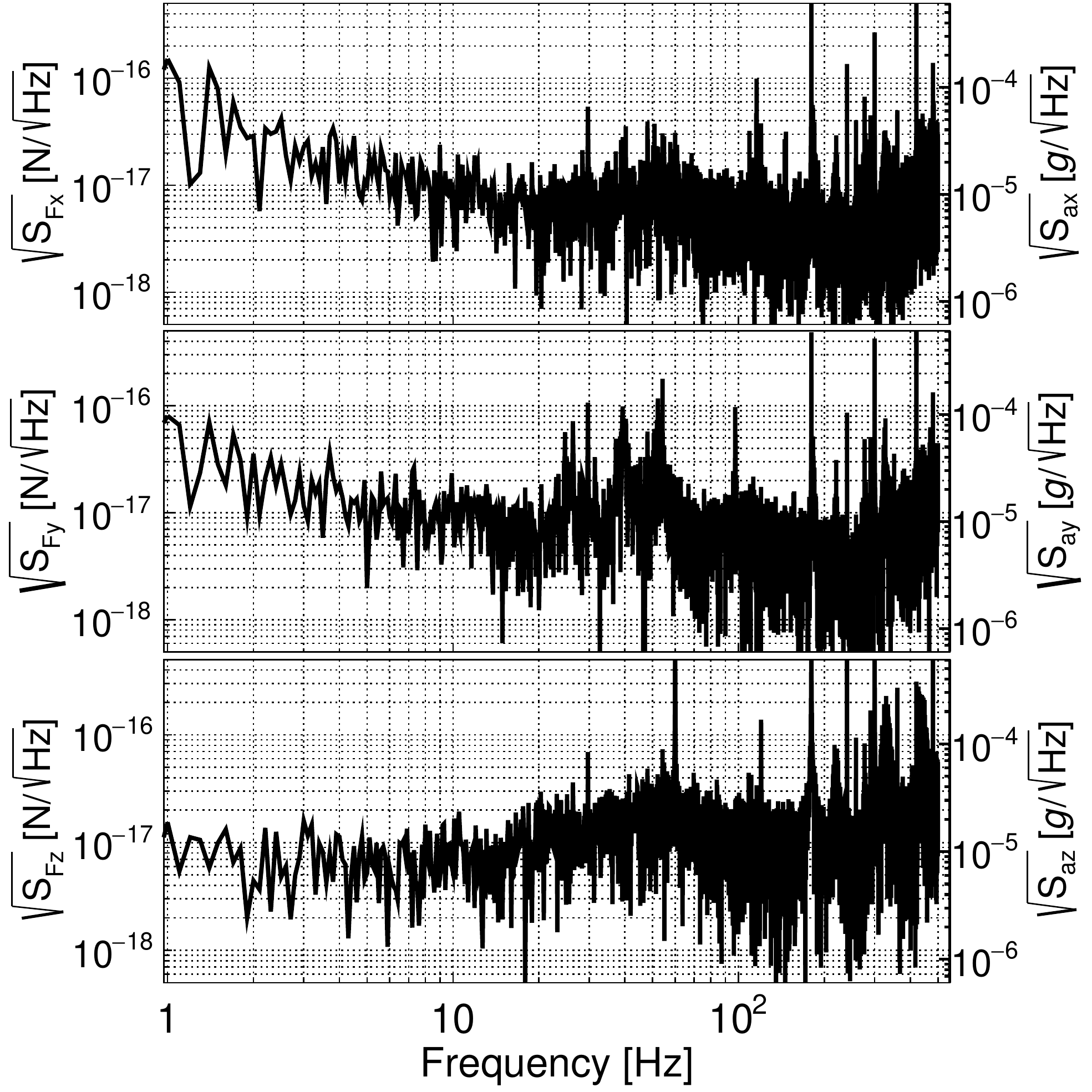}
    \caption{Force (left axis) and acceleration (right axis) sensitivity of the system using 4.7~$\upmu$m diameter MSs in the $x$ (upper panel), $y$ (middle panel), and $z$ (lower panel) directions, obtained by the procedure described in Section~\ref{SensitivitySection}. The power spectral density $S_{Fj}$ is estimated from the complex square of the Fourier transformation: $ \widetilde{F}_j^{\ast} \widetilde{F}_j $. }
    \label{ForceSensitivity}
\end{figure}

Utilizing the transfer function measurement described above, the noise spectrum of a neutral, 4.7~$\upmu$m diameter MS without any external driving field is converted into physical units (${\rm N}/\sqrt{\rm Hz}$), as shown in Fig.~\ref{ForceSensitivity}. This illustrates the typical force sensitivity of the system, reaching a level of $1\times10^{-17}~{\rm N}/\sqrt{\rm Hz}$. This force sensitivity is translated to an acceleration sensitivity of $12~\upmu g/\sqrt{\rm Hz}$, with $g=9.8~{\rm m/s}^2$. These values are comparable to the sensitivity previously reported \cite{PhysRevA.97.013842}. The observed noise floor is of additive, white Gaussian nature, and thus the sensitivity for any signal would improve with $\sqrt{t}$, with $t$ being the integration time, in the absence of background forces.

The expected Brownian motion of the MS at the vacuum level achieved is estimated to be $2 \times10^{-19}~{\rm N}/\sqrt{\rm Hz}$. In addition, it is demonstrated that the noise floor is the same at different pressures, below $\sim10^{-5}$~hPa. The shot noise is subdominant as well and estimated to be of the order of $3 \times10^{-19}~{\rm N}/\sqrt{\rm Hz}$ \cite{PhysRevA.97.013842}. Hence the observed noise floor is likely technical in nature, dominated by pointing fluctuations of the trapping and reference beams or due to electronics noise of the detection system. The improvement in low frequency $1/f$ noise as compared to Ref.~\cite{PhysRevA.97.013842} is mainly due to the enclosure of the free-space optics, resulting in reduced air currents and microphonics.

\section{Conclusion}
A system using a single optical tweezer to trap silica microspheres with diameters (masses) of 4.7 and 7.6~$\upmu$m (84 and 420~pg), intended to measure the force between the microsphere and another device in close proximity, is described. The small radial extent of the beam at the trap focus and the lack of significant non-Gaussian tails opens the possibility of bringing mechanical devices, such as thin silicon cantilevers, as close as 1.6~$\upmu$m from the surface of the microsphere. Electrodes surrounding the trap provide shielding, allow control over the translational degrees of freedom of charged microspheres and the rotational degrees of freedom of neutral microspheres, coupling to their electric dipole moment. The same electrode system allows for the calibration of the system's response to forces applied to the microsphere. The observed force sensitivity of $1\times10^{-17}$ ${\rm N}/\sqrt{\rm Hz}$ is limited by non-fundamental noise sources, so that further improvements are expected in the future. Auxiliary imaging of the optical trap and devices in its immediate vicinity from both the $y-z$ plane and the $x-y$ plane, allows precise alignment of devices relative to the trapped microsphere, making use of a number of nanopositioning stages. The system can be utilized for force measurements in the range of $1~\upmu$m-$10~\upmu$m, and for other applications requiring high precision, easily adjusted to trap larger or smaller microspheres.

\begin{acknowledgments}

This work was supported by NSF grant PHY1802952, ONR grant N00014-18-1-2409, and the Heising-Simons Foundation. A.K. acknowledges the partial support of a William~M. and Jane~D. Fairbank Postdoctoral Fellowship of Stanford University. N.P. acknowledges the partial support of the Koret Foundation. We acknowledge regular discussions on the physics of trapped microspheres with the group of Prof.~D.C.~Moore at Yale.  We also thank A.O.~Ames, A.D.~Rider, B.~Sandoval, J.~Singh, and T.~Yu, who contributed to early developments of the experimental apparatus and the Physics Machine Shop at Stanford for their technical support. The data that support the findings of this study are available from the corresponding author upon reasonable request.

\end{acknowledgments}

\bibliographystyle{apsrev4-1}
\bibliography{Instrumentation}

\begin{thebibliography}{53}%
\makeatletter
\providecommand \@ifxundefined [1]{%
 \@ifx{#1\undefined}
}%
\providecommand \@ifnum [1]{%
 \ifnum #1\expandafter \@firstoftwo
 \else \expandafter \@secondoftwo
 \fi
}%
\providecommand \@ifx [1]{%
 \ifx #1\expandafter \@firstoftwo
 \else \expandafter \@secondoftwo
 \fi
}%
\providecommand \natexlab [1]{#1}%
\providecommand \enquote  [1]{``#1''}%
\providecommand \bibnamefont  [1]{#1}%
\providecommand \bibfnamefont [1]{#1}%
\providecommand \citenamefont [1]{#1}%
\providecommand \href@noop [0]{\@secondoftwo}%
\providecommand \href [0]{\begingroup \@sanitize@url \@href}%
\providecommand \@href[1]{\@@startlink{#1}\@@href}%
\providecommand \@@href[1]{\endgroup#1\@@endlink}%
\providecommand \@sanitize@url [0]{\catcode `\\12\catcode `\$12\catcode
  `\&12\catcode `\#12\catcode `\^12\catcode `\_12\catcode `\%12\relax}%
\providecommand \@@startlink[1]{}%
\providecommand \@@endlink[0]{}%
\providecommand \url  [0]{\begingroup\@sanitize@url \@url }%
\providecommand \@url [1]{\endgroup\@href {#1}{\urlprefix }}%
\providecommand \urlprefix  [0]{URL }%
\providecommand \Eprint [0]{\href }%
\providecommand \doibase [0]{http://dx.doi.org/}%
\providecommand \selectlanguage [0]{\@gobble}%
\providecommand \bibinfo  [0]{\@secondoftwo}%
\providecommand \bibfield  [0]{\@secondoftwo}%
\providecommand \translation [1]{[#1]}%
\providecommand \BibitemOpen [0]{}%
\providecommand \bibitemStop [0]{}%
\providecommand \bibitemNoStop [0]{.\EOS\space}%
\providecommand \EOS [0]{\spacefactor3000\relax}%
\providecommand \BibitemShut  [1]{\csname bibitem#1\endcsname}%
\let\auto@bib@innerbib\@empty
\bibitem [{\citenamefont {Binnig}\ \emph {et~al.}(1986)\citenamefont {Binnig},
  \citenamefont {Quate},\ and\ \citenamefont {Gerber}}]{PhysRevLett.56.930}%
  \BibitemOpen
  \bibfield  {author} {\bibinfo {author} {\bibfnamefont {G.}~\bibnamefont
  {Binnig}}, \bibinfo {author} {\bibfnamefont {C.~F.}\ \bibnamefont {Quate}}, \
  and\ \bibinfo {author} {\bibfnamefont {C.}~\bibnamefont {Gerber}},\ }\href
  {\doibase 10.1103/PhysRevLett.56.930} {\bibfield  {journal} {\bibinfo
  {journal} {Phys. Rev. Lett.}\ }\textbf {\bibinfo {volume} {56}},\ \bibinfo
  {pages} {930} (\bibinfo {year} {1986})}\BibitemShut {NoStop}%
\bibitem [{\citenamefont {Abbott}\ \emph {et~al.}(2009)\citenamefont {Abbott}
  \emph {et~al.}}]{RepProgPhys.72.076901}%
  \BibitemOpen
  \bibfield  {author} {\bibinfo {author} {\bibfnamefont {B.~P.}\ \bibnamefont
  {Abbott}} \emph {et~al.},\ }\href {\doibase 10.1088/0034-4885/72/7/076901}
  {\bibfield  {journal} {\bibinfo  {journal} {Rep. Prog. Phys.}\ }\textbf
  {\bibinfo {volume} {72}},\ \bibinfo {pages} {076901} (\bibinfo {year}
  {2009})}\BibitemShut {NoStop}%
\bibitem [{\citenamefont {Aasi}\ \emph {et~al.}(2015)\citenamefont {Aasi} \emph
  {et~al.}}]{ClassQuantumGrav.32.074001}%
  \BibitemOpen
  \bibfield  {author} {\bibinfo {author} {\bibfnamefont {J.}~\bibnamefont
  {Aasi}} \emph {et~al.},\ }\href {\doibase 10.1088/0264-9381/32/7/074001}
  {\bibfield  {journal} {\bibinfo  {journal} {Class. Quantum Grav.}\ }\textbf
  {\bibinfo {volume} {32}},\ \bibinfo {pages} {074001} (\bibinfo {year}
  {2015})}\BibitemShut {NoStop}%
\bibitem [{\citenamefont {Acernese}\ \emph {et~al.}(2014)\citenamefont
  {Acernese} \emph {et~al.}}]{ClassQuantumGrav.32.024001}%
  \BibitemOpen
  \bibfield  {author} {\bibinfo {author} {\bibfnamefont {F.}~\bibnamefont
  {Acernese}} \emph {et~al.},\ }\href {\doibase 10.1088/0264-9381/32/2/024001}
  {\bibfield  {journal} {\bibinfo  {journal} {Class. Quantum Grav.}\ }\textbf
  {\bibinfo {volume} {32}},\ \bibinfo {pages} {024001} (\bibinfo {year}
  {2014})}\BibitemShut {NoStop}%
\bibitem [{\citenamefont {Akutsu}\ \emph {et~al.}(2019)\citenamefont {Akutsu}
  \emph {et~al.}}]{NatAstro.3.2397}%
  \BibitemOpen
  \bibfield  {author} {\bibinfo {author} {\bibfnamefont {T.}~\bibnamefont
  {Akutsu}} \emph {et~al.},\ }\href {\doibase 10.1038/s41550-018-0658-y}
  {\bibfield  {journal} {\bibinfo  {journal} {Nat. Astron.}\ }\textbf {\bibinfo
  {volume} {3}},\ \bibinfo {pages} {35} (\bibinfo {year} {2019})}\BibitemShut
  {NoStop}%
\bibitem [{\citenamefont {Peters}\ \emph {et~al.}(1999)\citenamefont {Peters},
  \citenamefont {Chung},\ and\ \citenamefont {Chu}}]{nature.400.849}%
  \BibitemOpen
  \bibfield  {author} {\bibinfo {author} {\bibfnamefont {A.}~\bibnamefont
  {Peters}}, \bibinfo {author} {\bibfnamefont {K.~Y.}\ \bibnamefont {Chung}}, \
  and\ \bibinfo {author} {\bibfnamefont {S.}~\bibnamefont {Chu}},\ }\href
  {\doibase 10.1038/23655} {\bibfield  {journal} {\bibinfo  {journal} {Nature}\
  }\textbf {\bibinfo {volume} {400}},\ \bibinfo {pages} {849} (\bibinfo {year}
  {1999})}\BibitemShut {NoStop}%
\bibitem [{\citenamefont {Peters}\ \emph {et~al.}(2001)\citenamefont {Peters},
  \citenamefont {Chung},\ and\ \citenamefont {Chu}}]{Metrologia.38.25}%
  \BibitemOpen
  \bibfield  {author} {\bibinfo {author} {\bibfnamefont {A.}~\bibnamefont
  {Peters}}, \bibinfo {author} {\bibfnamefont {K.~Y.}\ \bibnamefont {Chung}}, \
  and\ \bibinfo {author} {\bibfnamefont {S.}~\bibnamefont {Chu}},\ }\href
  {\doibase 10.1088/0026-1394/38/1/4} {\bibfield  {journal} {\bibinfo
  {journal} {Metrologia}\ }\textbf {\bibinfo {volume} {38}},\ \bibinfo {pages}
  {25} (\bibinfo {year} {2001})}\BibitemShut {NoStop}%
\bibitem [{\citenamefont {Biedermann}\ \emph {et~al.}(2015)\citenamefont
  {Biedermann}, \citenamefont {Wu}, \citenamefont {Deslauriers}, \citenamefont
  {Roy}, \citenamefont {Mahadeswaraswamy},\ and\ \citenamefont
  {Kasevich}}]{PhysRevA.91.033629}%
  \BibitemOpen
  \bibfield  {author} {\bibinfo {author} {\bibfnamefont {G.~W.}\ \bibnamefont
  {Biedermann}}, \bibinfo {author} {\bibfnamefont {X.}~\bibnamefont {Wu}},
  \bibinfo {author} {\bibfnamefont {L.}~\bibnamefont {Deslauriers}}, \bibinfo
  {author} {\bibfnamefont {S.}~\bibnamefont {Roy}}, \bibinfo {author}
  {\bibfnamefont {C.}~\bibnamefont {Mahadeswaraswamy}}, \ and\ \bibinfo
  {author} {\bibfnamefont {M.~A.}\ \bibnamefont {Kasevich}},\ }\href {\doibase
  10.1103/PhysRevA.91.033629} {\bibfield  {journal} {\bibinfo  {journal} {Phys.
  Rev. A}\ }\textbf {\bibinfo {volume} {91}},\ \bibinfo {pages} {033629}
  (\bibinfo {year} {2015})}\BibitemShut {NoStop}%
\bibitem [{\citenamefont {Hu}\ \emph {et~al.}(2013)\citenamefont {Hu},
  \citenamefont {Sun}, \citenamefont {Duan}, \citenamefont {Zhou},
  \citenamefont {Chen}, \citenamefont {Zhan}, \citenamefont {Zhang},\ and\
  \citenamefont {Luo}}]{PhysRevA.88.043610}%
  \BibitemOpen
  \bibfield  {author} {\bibinfo {author} {\bibfnamefont {Z.-K.}\ \bibnamefont
  {Hu}}, \bibinfo {author} {\bibfnamefont {B.-L.}\ \bibnamefont {Sun}},
  \bibinfo {author} {\bibfnamefont {X.-C.}\ \bibnamefont {Duan}}, \bibinfo
  {author} {\bibfnamefont {M.-K.}\ \bibnamefont {Zhou}}, \bibinfo {author}
  {\bibfnamefont {L.-L.}\ \bibnamefont {Chen}}, \bibinfo {author}
  {\bibfnamefont {S.}~\bibnamefont {Zhan}}, \bibinfo {author} {\bibfnamefont
  {Q.-Z.}\ \bibnamefont {Zhang}}, \ and\ \bibinfo {author} {\bibfnamefont
  {J.}~\bibnamefont {Luo}},\ }\href {\doibase 10.1103/PhysRevA.88.043610}
  {\bibfield  {journal} {\bibinfo  {journal} {Phys. Rev. A}\ }\textbf {\bibinfo
  {volume} {88}},\ \bibinfo {pages} {043610} (\bibinfo {year}
  {2013})}\BibitemShut {NoStop}%
\bibitem [{\citenamefont {Niebauer}\ \emph {et~al.}(1995)\citenamefont
  {Niebauer}, \citenamefont {Sasagawa}, \citenamefont {Faller}, \citenamefont
  {Hilt},\ and\ \citenamefont {Klopping}}]{Metrologia.32.159}%
  \BibitemOpen
  \bibfield  {author} {\bibinfo {author} {\bibfnamefont {T.~M.}\ \bibnamefont
  {Niebauer}}, \bibinfo {author} {\bibfnamefont {G.~S.}\ \bibnamefont
  {Sasagawa}}, \bibinfo {author} {\bibfnamefont {J.~E.}\ \bibnamefont
  {Faller}}, \bibinfo {author} {\bibfnamefont {R.}~\bibnamefont {Hilt}}, \ and\
  \bibinfo {author} {\bibfnamefont {F.}~\bibnamefont {Klopping}},\ }\href
  {\doibase 10.1088/0026-1394/32/3/004} {\bibfield  {journal} {\bibinfo
  {journal} {Metrologia}\ }\textbf {\bibinfo {volume} {32}},\ \bibinfo {pages}
  {159} (\bibinfo {year} {1995})}\BibitemShut {NoStop}%
\bibitem [{\citenamefont {Ashkin}(1970)}]{PhysRevLett.24.156}%
  \BibitemOpen
  \bibfield  {author} {\bibinfo {author} {\bibfnamefont {A.}~\bibnamefont
  {Ashkin}},\ }\href {\doibase 10.1103/PhysRevLett.24.156} {\bibfield
  {journal} {\bibinfo  {journal} {Phys. Rev. Lett.}\ }\textbf {\bibinfo
  {volume} {24}},\ \bibinfo {pages} {156} (\bibinfo {year} {1970})}\BibitemShut
  {NoStop}%
\bibitem [{\citenamefont {Ashkin}\ and\ \citenamefont
  {Dziedzic}(1971)}]{ApplPhysLett.19.283}%
  \BibitemOpen
  \bibfield  {author} {\bibinfo {author} {\bibfnamefont {A.}~\bibnamefont
  {Ashkin}}\ and\ \bibinfo {author} {\bibfnamefont {J.~M.}\ \bibnamefont
  {Dziedzic}},\ }\href {\doibase 10.1063/1.1653919} {\bibfield  {journal}
  {\bibinfo  {journal} {Appl. Phys. Lett.}\ }\textbf {\bibinfo {volume} {19}},\
  \bibinfo {pages} {283} (\bibinfo {year} {1971})}\BibitemShut {NoStop}%
\bibitem [{\citenamefont {Ashkin}\ and\ \citenamefont
  {Dziedzic}(1977)}]{ApplPhysLett.30.202}%
  \BibitemOpen
  \bibfield  {author} {\bibinfo {author} {\bibfnamefont {A.}~\bibnamefont
  {Ashkin}}\ and\ \bibinfo {author} {\bibfnamefont {J.~M.}\ \bibnamefont
  {Dziedzic}},\ }\href {\doibase 10.1063/1.89335} {\bibfield  {journal}
  {\bibinfo  {journal} {Appl. Phys. Lett.}\ }\textbf {\bibinfo {volume} {30}},\
  \bibinfo {pages} {202} (\bibinfo {year} {1977})}\BibitemShut {NoStop}%
\bibitem [{\citenamefont {{Ashkin}}\ \emph {et~al.}(1986)\citenamefont
  {{Ashkin}}, \citenamefont {{Dziedzic}}, \citenamefont {{Bjorkholm}},\ and\
  \citenamefont {{Chu}}}]{OptLett.11.288}%
  \BibitemOpen
  \bibfield  {author} {\bibinfo {author} {\bibfnamefont {A.}~\bibnamefont
  {{Ashkin}}}, \bibinfo {author} {\bibfnamefont {J.~M.}\ \bibnamefont
  {{Dziedzic}}}, \bibinfo {author} {\bibfnamefont {J.~E.}\ \bibnamefont
  {{Bjorkholm}}}, \ and\ \bibinfo {author} {\bibfnamefont {S.}~\bibnamefont
  {{Chu}}},\ }\href {\doibase 10.1364/OL.11.000288} {\bibfield  {journal}
  {\bibinfo  {journal} {Opt. Lett.}\ }\textbf {\bibinfo {volume} {11}},\
  \bibinfo {pages} {288} (\bibinfo {year} {1986})}\BibitemShut {NoStop}%
\bibitem [{\citenamefont {Fazal}\ and\ \citenamefont
  {Block}(2011)}]{NatPhot.5.318}%
  \BibitemOpen
  \bibfield  {author} {\bibinfo {author} {\bibfnamefont {F.~M.}\ \bibnamefont
  {Fazal}}\ and\ \bibinfo {author} {\bibfnamefont {S.~M.}\ \bibnamefont
  {Block}},\ }\href {\doibase 10.1038/nphoton.2011.100} {\bibfield  {journal}
  {\bibinfo  {journal} {Nat. Photon.}\ }\textbf {\bibinfo {volume} {5}},\
  \bibinfo {pages} {318} (\bibinfo {year} {2011})}\BibitemShut {NoStop}%
\bibitem [{\citenamefont {Gutiérrez-Medina}\ \emph {et~al.}(2010)\citenamefont
  {Gutiérrez-Medina}, \citenamefont {Andreasson}, \citenamefont {Greenleaf},
  \citenamefont {LaPorta},\ and\ \citenamefont
  {Block}}]{MethEnzymology.475.377}%
  \BibitemOpen
  \bibfield  {author} {\bibinfo {author} {\bibfnamefont {B.}~\bibnamefont
  {Gutiérrez-Medina}}, \bibinfo {author} {\bibfnamefont {J.~O.}\ \bibnamefont
  {Andreasson}}, \bibinfo {author} {\bibfnamefont {W.~J.}\ \bibnamefont
  {Greenleaf}}, \bibinfo {author} {\bibfnamefont {A.}~\bibnamefont {LaPorta}},
  \ and\ \bibinfo {author} {\bibfnamefont {S.~M.}\ \bibnamefont {Block}},\
  }\href {\doibase https://doi.org/10.1016/S0076-6879(10)75015-1} {\ \bibinfo
  {series} {Methods Enzymol.},\ \textbf {\bibinfo {volume} {475}},\ \bibinfo
  {pages} {377 } (\bibinfo {year} {2010})}\BibitemShut {NoStop}%
\bibitem [{\citenamefont {Neuman}\ and\ \citenamefont
  {Block}(2004)}]{RevSciInstrum.75.2787}%
  \BibitemOpen
  \bibfield  {author} {\bibinfo {author} {\bibfnamefont {K.~C.}\ \bibnamefont
  {Neuman}}\ and\ \bibinfo {author} {\bibfnamefont {S.~M.}\ \bibnamefont
  {Block}},\ }\href {\doibase 10.1063/1.1785844} {\bibfield  {journal}
  {\bibinfo  {journal} {Rev. Sci. Instrum.}\ }\textbf {\bibinfo {volume}
  {75}},\ \bibinfo {pages} {2787} (\bibinfo {year} {2004})}\BibitemShut
  {NoStop}%
\bibitem [{\citenamefont {Perkins}\ \emph {et~al.}(1994)\citenamefont
  {Perkins}, \citenamefont {Smith},\ and\ \citenamefont
  {Chu}}]{Science.264.819}%
  \BibitemOpen
  \bibfield  {author} {\bibinfo {author} {\bibfnamefont {T.}~\bibnamefont
  {Perkins}}, \bibinfo {author} {\bibfnamefont {D.}~\bibnamefont {Smith}}, \
  and\ \bibinfo {author} {\bibfnamefont {S.}~\bibnamefont {Chu}},\ }\href
  {\doibase 10.1126/science.8171335} {\bibfield  {journal} {\bibinfo  {journal}
  {Science}\ }\textbf {\bibinfo {volume} {264}},\ \bibinfo {pages} {819}
  (\bibinfo {year} {1994})}\BibitemShut {NoStop}%
\bibitem [{\citenamefont {Li}\ \emph {et~al.}(2011)\citenamefont {Li},
  \citenamefont {Kheifets},\ and\ \citenamefont {Raizen}}]{NatPhys.7.527}%
  \BibitemOpen
  \bibfield  {author} {\bibinfo {author} {\bibfnamefont {T.}~\bibnamefont
  {Li}}, \bibinfo {author} {\bibfnamefont {S.}~\bibnamefont {Kheifets}}, \ and\
  \bibinfo {author} {\bibfnamefont {M.~G.}\ \bibnamefont {Raizen}},\ }\href
  {\doibase 10.1038/nphys1952} {\bibfield  {journal} {\bibinfo  {journal} {Nat.
  Phys.}\ }\textbf {\bibinfo {volume} {7}},\ \bibinfo {pages} {527} (\bibinfo
  {year} {2011})}\BibitemShut {NoStop}%
\bibitem [{\citenamefont {Ranjit}\ \emph {et~al.}(2016)\citenamefont {Ranjit},
  \citenamefont {Cunningham}, \citenamefont {Casey},\ and\ \citenamefont
  {Geraci}}]{PhysRevA.93.053801}%
  \BibitemOpen
  \bibfield  {author} {\bibinfo {author} {\bibfnamefont {G.}~\bibnamefont
  {Ranjit}}, \bibinfo {author} {\bibfnamefont {M.}~\bibnamefont {Cunningham}},
  \bibinfo {author} {\bibfnamefont {K.}~\bibnamefont {Casey}}, \ and\ \bibinfo
  {author} {\bibfnamefont {A.~A.}\ \bibnamefont {Geraci}},\ }\href {\doibase
  10.1103/PhysRevA.93.053801} {\bibfield  {journal} {\bibinfo  {journal} {Phys.
  Rev. A}\ }\textbf {\bibinfo {volume} {93}},\ \bibinfo {pages} {053801}
  (\bibinfo {year} {2016})}\BibitemShut {NoStop}%
\bibitem [{\citenamefont {Hempston}\ \emph {et~al.}(2017)\citenamefont
  {Hempston}, \citenamefont {Vovrosh}, \citenamefont {Toroš}, \citenamefont
  {Winstone}, \citenamefont {Rashid},\ and\ \citenamefont
  {Ulbricht}}]{ApplPhysLett.111.133111}%
  \BibitemOpen
  \bibfield  {author} {\bibinfo {author} {\bibfnamefont {D.}~\bibnamefont
  {Hempston}}, \bibinfo {author} {\bibfnamefont {J.}~\bibnamefont {Vovrosh}},
  \bibinfo {author} {\bibfnamefont {M.}~\bibnamefont {Toroš}}, \bibinfo
  {author} {\bibfnamefont {G.}~\bibnamefont {Winstone}}, \bibinfo {author}
  {\bibfnamefont {M.}~\bibnamefont {Rashid}}, \ and\ \bibinfo {author}
  {\bibfnamefont {H.}~\bibnamefont {Ulbricht}},\ }\href {\doibase
  10.1063/1.4993555} {\bibfield  {journal} {\bibinfo  {journal} {Appl. Phys.
  Lett.}\ }\textbf {\bibinfo {volume} {111}},\ \bibinfo {pages} {133111}
  (\bibinfo {year} {2017})}\BibitemShut {NoStop}%
\bibitem [{\citenamefont {Blakemore}\ \emph
  {et~al.}(2019{\natexlab{a}})\citenamefont {Blakemore}, \citenamefont {Rider},
  \citenamefont {Roy}, \citenamefont {Wang}, \citenamefont {Kawasaki},\ and\
  \citenamefont {Gratta}}]{PhysRevA.99.023816}%
  \BibitemOpen
  \bibfield  {author} {\bibinfo {author} {\bibfnamefont {C.~P.}\ \bibnamefont
  {Blakemore}}, \bibinfo {author} {\bibfnamefont {A.~D.}\ \bibnamefont
  {Rider}}, \bibinfo {author} {\bibfnamefont {S.}~\bibnamefont {Roy}}, \bibinfo
  {author} {\bibfnamefont {Q.}~\bibnamefont {Wang}}, \bibinfo {author}
  {\bibfnamefont {A.}~\bibnamefont {Kawasaki}}, \ and\ \bibinfo {author}
  {\bibfnamefont {G.}~\bibnamefont {Gratta}},\ }\href {\doibase
  10.1103/PhysRevA.99.023816} {\bibfield  {journal} {\bibinfo  {journal} {Phys.
  Rev. A}\ }\textbf {\bibinfo {volume} {99}},\ \bibinfo {pages} {023816}
  (\bibinfo {year} {2019}{\natexlab{a}})}\BibitemShut {NoStop}%
\bibitem [{\citenamefont {Gieseler}\ \emph {et~al.}(2012)\citenamefont
  {Gieseler}, \citenamefont {Deutsch}, \citenamefont {Quidant},\ and\
  \citenamefont {Novotny}}]{PhysRevLett.109.103603}%
  \BibitemOpen
  \bibfield  {author} {\bibinfo {author} {\bibfnamefont {J.}~\bibnamefont
  {Gieseler}}, \bibinfo {author} {\bibfnamefont {B.}~\bibnamefont {Deutsch}},
  \bibinfo {author} {\bibfnamefont {R.}~\bibnamefont {Quidant}}, \ and\
  \bibinfo {author} {\bibfnamefont {L.}~\bibnamefont {Novotny}},\ }\href
  {\doibase 10.1103/PhysRevLett.109.103603} {\bibfield  {journal} {\bibinfo
  {journal} {Phys. Rev. Lett.}\ }\textbf {\bibinfo {volume} {109}},\ \bibinfo
  {pages} {103603} (\bibinfo {year} {2012})}\BibitemShut {NoStop}%
\bibitem [{\citenamefont {Yin}\ \emph {et~al.}(2013)\citenamefont {Yin},
  \citenamefont {Geraci},\ and\ \citenamefont {Li}}]{IntJModPhys.B27.1330018}%
  \BibitemOpen
  \bibfield  {author} {\bibinfo {author} {\bibfnamefont {Z.~Q.}\ \bibnamefont
  {Yin}}, \bibinfo {author} {\bibfnamefont {A.~A.}\ \bibnamefont {Geraci}}, \
  and\ \bibinfo {author} {\bibfnamefont {T.}~\bibnamefont {Li}},\ }\href
  {\doibase 10.1142/S0217979213300181} {\bibfield  {journal} {\bibinfo
  {journal} {Int. J. Mod. Phys.}\ }\textbf {\bibinfo {volume} {B27}},\ \bibinfo
  {pages} {1330018} (\bibinfo {year} {2013})}\BibitemShut {NoStop}%
\bibitem [{\citenamefont {Gieseler}\ \emph {et~al.}(2013)\citenamefont
  {Gieseler}, \citenamefont {Novotny},\ and\ \citenamefont
  {Quidant}}]{NatPhys.12.806}%
  \BibitemOpen
  \bibfield  {author} {\bibinfo {author} {\bibfnamefont {J.}~\bibnamefont
  {Gieseler}}, \bibinfo {author} {\bibfnamefont {L.}~\bibnamefont {Novotny}}, \
  and\ \bibinfo {author} {\bibfnamefont {R.}~\bibnamefont {Quidant}},\ }\href
  {\doibase 10.1038/nphys2798} {\bibfield  {journal} {\bibinfo  {journal} {Nat.
  Phys.}\ }\textbf {\bibinfo {volume} {9}},\ \bibinfo {pages} {806} (\bibinfo
  {year} {2013})}\BibitemShut {NoStop}%
\bibitem [{\citenamefont {Monteiro}\ \emph {et~al.}(2020)\citenamefont
  {Monteiro}, \citenamefont {Li}, \citenamefont {Afek}, \citenamefont {Li},
  \citenamefont {Mossman},\ and\ \citenamefont {Moore}}]{2001.1093}%
  \BibitemOpen
  \bibfield  {author} {\bibinfo {author} {\bibfnamefont {F.}~\bibnamefont
  {Monteiro}}, \bibinfo {author} {\bibfnamefont {W.}~\bibnamefont {Li}},
  \bibinfo {author} {\bibfnamefont {G.}~\bibnamefont {Afek}}, \bibinfo {author}
  {\bibfnamefont {C.-l.}\ \bibnamefont {Li}}, \bibinfo {author} {\bibfnamefont
  {M.}~\bibnamefont {Mossman}}, \ and\ \bibinfo {author} {\bibfnamefont
  {D.~C.}\ \bibnamefont {Moore}},\ }\href@noop {} {} (\bibinfo {year} {2020}),\
  \Eprint {http://arxiv.org/abs/2001.10931} {arXiv:2001.10931 [physics.optics]}
  \BibitemShut {NoStop}%
\bibitem [{\citenamefont {Ahn}\ \emph {et~al.}(2020)\citenamefont {Ahn},
  \citenamefont {Xu}, \citenamefont {Bang}, \citenamefont {Ju}, \citenamefont
  {Gao},\ and\ \citenamefont {Li}}]{NatNanotechnol.2.89}%
  \BibitemOpen
  \bibfield  {author} {\bibinfo {author} {\bibfnamefont {J.}~\bibnamefont
  {Ahn}}, \bibinfo {author} {\bibfnamefont {Z.}~\bibnamefont {Xu}}, \bibinfo
  {author} {\bibfnamefont {J.}~\bibnamefont {Bang}}, \bibinfo {author}
  {\bibfnamefont {P.}~\bibnamefont {Ju}}, \bibinfo {author} {\bibfnamefont
  {X.}~\bibnamefont {Gao}}, \ and\ \bibinfo {author} {\bibfnamefont
  {T.}~\bibnamefont {Li}},\ }\href {\doibase 10.1038/s41565-019-0605-9}
  {\bibfield  {journal} {\bibinfo  {journal} {Nat. Nanotechnol.}\ }\textbf
  {\bibinfo {volume} {15}},\ \bibinfo {pages} {89} (\bibinfo {year}
  {2020})}\BibitemShut {NoStop}%
\bibitem [{\citenamefont {Deli{\'c}}\ \emph {et~al.}(2020)\citenamefont
  {Deli{\'c}}, \citenamefont {Reisenbauer}, \citenamefont {Dare}, \citenamefont
  {Grass}, \citenamefont {Vuleti{\'c}}, \citenamefont {Kiesel},\ and\
  \citenamefont {Aspelmeyer}}]{Science.367.6480}%
  \BibitemOpen
  \bibfield  {author} {\bibinfo {author} {\bibfnamefont {U.}~\bibnamefont
  {Deli{\'c}}}, \bibinfo {author} {\bibfnamefont {M.}~\bibnamefont
  {Reisenbauer}}, \bibinfo {author} {\bibfnamefont {K.}~\bibnamefont {Dare}},
  \bibinfo {author} {\bibfnamefont {D.}~\bibnamefont {Grass}}, \bibinfo
  {author} {\bibfnamefont {V.}~\bibnamefont {Vuleti{\'c}}}, \bibinfo {author}
  {\bibfnamefont {N.}~\bibnamefont {Kiesel}}, \ and\ \bibinfo {author}
  {\bibfnamefont {M.}~\bibnamefont {Aspelmeyer}},\ }\href {\doibase
  10.1126/science.aba3993} {\bibfield  {journal} {\bibinfo  {journal}
  {Science}\ }\textbf {\bibinfo {volume} {367}},\ \bibinfo {pages} {892}
  (\bibinfo {year} {2020})}\BibitemShut {NoStop}%
\bibitem [{\citenamefont {Rider}\ \emph {et~al.}(2016)\citenamefont {Rider},
  \citenamefont {Moore}, \citenamefont {Blakemore}, \citenamefont {Louis},
  \citenamefont {Lu},\ and\ \citenamefont {Gratta}}]{PhysRevLett.117.101101}%
  \BibitemOpen
  \bibfield  {author} {\bibinfo {author} {\bibfnamefont {A.~D.}\ \bibnamefont
  {Rider}}, \bibinfo {author} {\bibfnamefont {D.~C.}\ \bibnamefont {Moore}},
  \bibinfo {author} {\bibfnamefont {C.~P.}\ \bibnamefont {Blakemore}}, \bibinfo
  {author} {\bibfnamefont {M.}~\bibnamefont {Louis}}, \bibinfo {author}
  {\bibfnamefont {M.}~\bibnamefont {Lu}}, \ and\ \bibinfo {author}
  {\bibfnamefont {G.}~\bibnamefont {Gratta}},\ }\href {\doibase
  10.1103/PhysRevLett.117.101101} {\bibfield  {journal} {\bibinfo  {journal}
  {Phys. Rev. Lett.}\ }\textbf {\bibinfo {volume} {117}},\ \bibinfo {pages}
  {101101} (\bibinfo {year} {2016})}\BibitemShut {NoStop}%
\bibitem [{\citenamefont {Winstone}\ \emph {et~al.}(2018)\citenamefont
  {Winstone}, \citenamefont {Bennett}, \citenamefont {Rademacher},
  \citenamefont {Rashid}, \citenamefont {Buhmann},\ and\ \citenamefont
  {Ulbricht}}]{PhysRevA.98.053831}%
  \BibitemOpen
  \bibfield  {author} {\bibinfo {author} {\bibfnamefont {G.}~\bibnamefont
  {Winstone}}, \bibinfo {author} {\bibfnamefont {R.}~\bibnamefont {Bennett}},
  \bibinfo {author} {\bibfnamefont {M.}~\bibnamefont {Rademacher}}, \bibinfo
  {author} {\bibfnamefont {M.}~\bibnamefont {Rashid}}, \bibinfo {author}
  {\bibfnamefont {S.}~\bibnamefont {Buhmann}}, \ and\ \bibinfo {author}
  {\bibfnamefont {H.}~\bibnamefont {Ulbricht}},\ }\href {\doibase
  10.1103/PhysRevA.98.053831} {\bibfield  {journal} {\bibinfo  {journal} {Phys.
  Rev. A}\ }\textbf {\bibinfo {volume} {98}},\ \bibinfo {pages} {053831}
  (\bibinfo {year} {2018})}\BibitemShut {NoStop}%
\bibitem [{\citenamefont {Diehl}\ \emph {et~al.}(2018)\citenamefont {Diehl},
  \citenamefont {Hebestreit}, \citenamefont {Reimann}, \citenamefont
  {Tebbenjohanns}, \citenamefont {Frimmer},\ and\ \citenamefont
  {Novotny}}]{PhysRevA.98.013852}%
  \BibitemOpen
  \bibfield  {author} {\bibinfo {author} {\bibfnamefont {R.}~\bibnamefont
  {Diehl}}, \bibinfo {author} {\bibfnamefont {E.}~\bibnamefont {Hebestreit}},
  \bibinfo {author} {\bibfnamefont {R.}~\bibnamefont {Reimann}}, \bibinfo
  {author} {\bibfnamefont {F.}~\bibnamefont {Tebbenjohanns}}, \bibinfo {author}
  {\bibfnamefont {M.}~\bibnamefont {Frimmer}}, \ and\ \bibinfo {author}
  {\bibfnamefont {L.}~\bibnamefont {Novotny}},\ }\href {\doibase
  10.1103/PhysRevA.98.013851} {\bibfield  {journal} {\bibinfo  {journal} {Phys.
  Rev. A}\ }\textbf {\bibinfo {volume} {98}},\ \bibinfo {pages} {013851}
  (\bibinfo {year} {2018})}\BibitemShut {NoStop}%
\bibitem [{\citenamefont {Rider}\ \emph {et~al.}(2018)\citenamefont {Rider},
  \citenamefont {Blakemore}, \citenamefont {Gratta},\ and\ \citenamefont
  {Moore}}]{PhysRevA.97.013842}%
  \BibitemOpen
  \bibfield  {author} {\bibinfo {author} {\bibfnamefont {A.~D.}\ \bibnamefont
  {Rider}}, \bibinfo {author} {\bibfnamefont {C.~P.}\ \bibnamefont
  {Blakemore}}, \bibinfo {author} {\bibfnamefont {G.}~\bibnamefont {Gratta}}, \
  and\ \bibinfo {author} {\bibfnamefont {D.~C.}\ \bibnamefont {Moore}},\ }\href
  {\doibase 10.1103/PhysRevA.97.013842} {\bibfield  {journal} {\bibinfo
  {journal} {Phys. Rev. A}\ }\textbf {\bibinfo {volume} {97}},\ \bibinfo
  {pages} {013842} (\bibinfo {year} {2018})}\BibitemShut {NoStop}%
\bibitem [{\citenamefont {Rider}\ \emph {et~al.}(2019)\citenamefont {Rider},
  \citenamefont {Blakemore}, \citenamefont {Kawasaki}, \citenamefont {Priel},
  \citenamefont {Roy},\ and\ \citenamefont {Gratta}}]{PhysRevA.99.041802}%
  \BibitemOpen
  \bibfield  {author} {\bibinfo {author} {\bibfnamefont {A.~D.}\ \bibnamefont
  {Rider}}, \bibinfo {author} {\bibfnamefont {C.~P.}\ \bibnamefont
  {Blakemore}}, \bibinfo {author} {\bibfnamefont {A.}~\bibnamefont {Kawasaki}},
  \bibinfo {author} {\bibfnamefont {N.}~\bibnamefont {Priel}}, \bibinfo
  {author} {\bibfnamefont {S.}~\bibnamefont {Roy}}, \ and\ \bibinfo {author}
  {\bibfnamefont {G.}~\bibnamefont {Gratta}},\ }\href {\doibase
  10.1103/PhysRevA.99.041802} {\bibfield  {journal} {\bibinfo  {journal} {Phys.
  Rev. A}\ }\textbf {\bibinfo {volume} {99}},\ \bibinfo {pages} {041802(R)}
  (\bibinfo {year} {2019})}\BibitemShut {NoStop}%
\bibitem [{\citenamefont {Geraci}\ \emph {et~al.}(2010)\citenamefont {Geraci},
  \citenamefont {Papp},\ and\ \citenamefont
  {Kitching}}]{PhysRevLett.105.101101}%
  \BibitemOpen
  \bibfield  {author} {\bibinfo {author} {\bibfnamefont {A.~A.}\ \bibnamefont
  {Geraci}}, \bibinfo {author} {\bibfnamefont {S.~B.}\ \bibnamefont {Papp}}, \
  and\ \bibinfo {author} {\bibfnamefont {J.}~\bibnamefont {Kitching}},\ }\href
  {\doibase 10.1103/PhysRevLett.105.101101} {\bibfield  {journal} {\bibinfo
  {journal} {Phys. Rev. Lett.}\ }\textbf {\bibinfo {volume} {105}},\ \bibinfo
  {pages} {101101} (\bibinfo {year} {2010})}\BibitemShut {NoStop}%
\bibitem [{\citenamefont {Casimir}(1948)}]{Casimir}%
  \BibitemOpen
  \bibfield  {author} {\bibinfo {author} {\bibfnamefont {H.~B.~G.}\
  \bibnamefont {Casimir}},\ }\href@noop {} {\bibfield  {journal} {\bibinfo
  {journal} {Proceedings of the Royal Netherlands Academy of Arts and
  Sciences}\ }\textbf {\bibinfo {volume} {51}},\ \bibinfo {pages} {793}
  (\bibinfo {year} {1948})}\BibitemShut {NoStop}%
\bibitem [{\citenamefont {Lamoreaux}(1997)}]{PhysRevLett.78.5}%
  \BibitemOpen
  \bibfield  {author} {\bibinfo {author} {\bibfnamefont {S.~K.}\ \bibnamefont
  {Lamoreaux}},\ }\href {\doibase 10.1103/PhysRevLett.78.5} {\bibfield
  {journal} {\bibinfo  {journal} {Phys. Rev. Lett.}\ }\textbf {\bibinfo
  {volume} {78}},\ \bibinfo {pages} {5} (\bibinfo {year} {1997})}\BibitemShut
  {NoStop}%
\bibitem [{\citenamefont {Mohideen}\ and\ \citenamefont
  {Roy}(1998)}]{PhysRevLett.81.4549}%
  \BibitemOpen
  \bibfield  {author} {\bibinfo {author} {\bibfnamefont {U.}~\bibnamefont
  {Mohideen}}\ and\ \bibinfo {author} {\bibfnamefont {A.}~\bibnamefont {Roy}},\
  }\href {\doibase 10.1103/PhysRevLett.81.4549} {\bibfield  {journal} {\bibinfo
   {journal} {Phys. Rev. Lett.}\ }\textbf {\bibinfo {volume} {81}},\ \bibinfo
  {pages} {4549} (\bibinfo {year} {1998})}\BibitemShut {NoStop}%
\bibitem [{\citenamefont {Ederth}(2000)}]{PhysRevA.62.062104}%
  \BibitemOpen
  \bibfield  {author} {\bibinfo {author} {\bibfnamefont {T.}~\bibnamefont
  {Ederth}},\ }\href {\doibase 10.1103/PhysRevA.62.062104} {\bibfield
  {journal} {\bibinfo  {journal} {Phys. Rev. A}\ }\textbf {\bibinfo {volume}
  {62}},\ \bibinfo {pages} {062104} (\bibinfo {year} {2000})}\BibitemShut
  {NoStop}%
\bibitem [{\citenamefont {Bressi}\ \emph {et~al.}(2002)\citenamefont {Bressi},
  \citenamefont {Carugno}, \citenamefont {Onofrio},\ and\ \citenamefont
  {Ruoso}}]{PhysRevLett.88.041804}%
  \BibitemOpen
  \bibfield  {author} {\bibinfo {author} {\bibfnamefont {G.}~\bibnamefont
  {Bressi}}, \bibinfo {author} {\bibfnamefont {G.}~\bibnamefont {Carugno}},
  \bibinfo {author} {\bibfnamefont {R.}~\bibnamefont {Onofrio}}, \ and\
  \bibinfo {author} {\bibfnamefont {G.}~\bibnamefont {Ruoso}},\ }\href
  {\doibase 10.1103/PhysRevLett.88.041804} {\bibfield  {journal} {\bibinfo
  {journal} {Phys. Rev. Lett.}\ }\textbf {\bibinfo {volume} {88}},\ \bibinfo
  {pages} {041804} (\bibinfo {year} {2002})}\BibitemShut {NoStop}%
\bibitem [{\citenamefont {Decca}\ \emph {et~al.}(2003)\citenamefont {Decca},
  \citenamefont {L\'opez}, \citenamefont {Fischbach},\ and\ \citenamefont
  {Krause}}]{PhysRevLett.91.050402}%
  \BibitemOpen
  \bibfield  {author} {\bibinfo {author} {\bibfnamefont {R.~S.}\ \bibnamefont
  {Decca}}, \bibinfo {author} {\bibfnamefont {D.}~\bibnamefont {L\'opez}},
  \bibinfo {author} {\bibfnamefont {E.}~\bibnamefont {Fischbach}}, \ and\
  \bibinfo {author} {\bibfnamefont {D.~E.}\ \bibnamefont {Krause}},\ }\href
  {\doibase 10.1103/PhysRevLett.91.050402} {\bibfield  {journal} {\bibinfo
  {journal} {Phys. Rev. Lett.}\ }\textbf {\bibinfo {volume} {91}},\ \bibinfo
  {pages} {050402} (\bibinfo {year} {2003})}\BibitemShut {NoStop}%
\bibitem [{\citenamefont {Ether}\ \emph {et~al.}(2015)\citenamefont {Ether}
  \emph {et~al.}}]{EPL.112.44001}%
  \BibitemOpen
  \bibfield  {author} {\bibinfo {author} {\bibfnamefont {D.~S.}\ \bibnamefont
  {Ether}} \emph {et~al.},\ }\href {\doibase 10.1209/0295-5075/112/44001}
  {\bibfield  {journal} {\bibinfo  {journal} {Europhys. Lett.}\ }\textbf
  {\bibinfo {volume} {112}},\ \bibinfo {pages} {44001} (\bibinfo {year}
  {2015})}\BibitemShut {NoStop}%
\bibitem [{\citenamefont {Arvanitaki}\ and\ \citenamefont
  {Geraci}(2013)}]{PhysRevLett.110.071105}%
  \BibitemOpen
  \bibfield  {author} {\bibinfo {author} {\bibfnamefont {A.}~\bibnamefont
  {Arvanitaki}}\ and\ \bibinfo {author} {\bibfnamefont {A.~A.}\ \bibnamefont
  {Geraci}},\ }\href {\doibase 10.1103/PhysRevLett.110.071105} {\bibfield
  {journal} {\bibinfo  {journal} {Phys. Rev. Lett.}\ }\textbf {\bibinfo
  {volume} {110}},\ \bibinfo {pages} {071105} (\bibinfo {year}
  {2013})}\BibitemShut {NoStop}%
\bibitem [{\citenamefont {Kawasaki}(2020)}]{ClassQuantumGrav.37.075002}%
  \BibitemOpen
  \bibfield  {author} {\bibinfo {author} {\bibfnamefont {A.}~\bibnamefont
  {Kawasaki}},\ }\href {\doibase 10.1088/1361-6382/ab6f80} {\bibfield
  {journal} {\bibinfo  {journal} {Class. Quantum Grav.}\ }\textbf {\bibinfo
  {volume} {37}},\ \bibinfo {pages} {075002} (\bibinfo {year}
  {2020})}\BibitemShut {NoStop}%
\bibitem [{\citenamefont {Kawasaki}(2019)}]{PhysRevD.99.023005}%
  \BibitemOpen
  \bibfield  {author} {\bibinfo {author} {\bibfnamefont {A.}~\bibnamefont
  {Kawasaki}},\ }\href {\doibase 10.1103/PhysRevD.99.023005} {\bibfield
  {journal} {\bibinfo  {journal} {Phys. Rev. D}\ }\textbf {\bibinfo {volume}
  {99}},\ \bibinfo {pages} {023005} (\bibinfo {year} {2019})}\BibitemShut
  {NoStop}%
\bibitem [{\citenamefont {Blakemore}\ \emph {et~al.}(2020)\citenamefont
  {Blakemore}, \citenamefont {Martin}, \citenamefont {Fieguth}, \citenamefont
  {Kawasaki}, \citenamefont {Priel}, \citenamefont {Rider},\ and\ \citenamefont
  {Gratta}}]{JVSTB.38.024201}%
  \BibitemOpen
  \bibfield  {author} {\bibinfo {author} {\bibfnamefont {C.~P.}\ \bibnamefont
  {Blakemore}}, \bibinfo {author} {\bibfnamefont {D.}~\bibnamefont {Martin}},
  \bibinfo {author} {\bibfnamefont {A.}~\bibnamefont {Fieguth}}, \bibinfo
  {author} {\bibfnamefont {A.~k.}\ \bibnamefont {Kawasaki}}, \bibinfo {author}
  {\bibfnamefont {N.}~\bibnamefont {Priel}}, \bibinfo {author} {\bibfnamefont
  {A.~D.}\ \bibnamefont {Rider}}, \ and\ \bibinfo {author} {\bibfnamefont
  {G.}~\bibnamefont {Gratta}},\ }\href {\doibase 10.1116/1.5139638} {\bibfield
  {journal} {\bibinfo  {journal} {J. Vac. Sci. Technol. B}\ }\textbf {\bibinfo
  {volume} {38}},\ \bibinfo {pages} {024201} (\bibinfo {year}
  {2020})}\BibitemShut {NoStop}%
\bibitem [{\citenamefont {{Wang}}\ \emph {et~al.}(2017)\citenamefont {{Wang}},
  \citenamefont {{Rider}}, \citenamefont {{Moore}}, \citenamefont
  {{Blakemore}}, \citenamefont {{Cao}},\ and\ \citenamefont
  {{Gratta}}}]{AttractorPaper}%
  \BibitemOpen
  \bibfield  {author} {\bibinfo {author} {\bibfnamefont {Q.}~\bibnamefont
  {{Wang}}}, \bibinfo {author} {\bibfnamefont {A.~D.}\ \bibnamefont {{Rider}}},
  \bibinfo {author} {\bibfnamefont {D.~C.}\ \bibnamefont {{Moore}}}, \bibinfo
  {author} {\bibfnamefont {C.~P.}\ \bibnamefont {{Blakemore}}}, \bibinfo
  {author} {\bibfnamefont {L.}~\bibnamefont {{Cao}}}, \ and\ \bibinfo {author}
  {\bibfnamefont {G.}~\bibnamefont {{Gratta}}},\ }in\ \href {\doibase
  10.1109/ECTC.2017.274} {\emph {\bibinfo {booktitle} {2017 IEEE 67th Electron.
  Compon. and Technol. Conf. (ECTC)}}}\ (\bibinfo {year} {2017})\ pp.\ \bibinfo
  {pages} {1773--1778}\BibitemShut {NoStop}%
\bibitem [{\citenamefont {Li}(2011)}]{TLiThesis}%
  \BibitemOpen
  \bibfield  {author} {\bibinfo {author} {\bibfnamefont {T.}~\bibnamefont
  {Li}},\ }\href@noop {} {\bibfield  {journal} {\bibinfo  {journal} {Ph.D.
  Thesis, The University of Texas, Austin}\ } (\bibinfo {year}
  {2011})}\BibitemShut {NoStop}%
\bibitem [{\citenamefont {\relax Bangs~Laboratories}()}]{bangs_laboratories}%
  \BibitemOpen
  \bibfield  {author} {\bibinfo {author} {\bibnamefont {\relax
  Bangs~Laboratories}},\ }\href@noop {} {}\bibinfo {howpublished}
  {[\url{http://www.bangslabs.com/}]}\BibitemShut {NoStop}%
\bibitem [{\citenamefont {Blakemore}\ \emph
  {et~al.}(2019{\natexlab{b}})\citenamefont {Blakemore}, \citenamefont {Rider},
  \citenamefont {Roy}, \citenamefont {Fieguth}, \citenamefont {Kawasaki},
  \citenamefont {Priel},\ and\ \citenamefont
  {Gratta}}]{PhysRevApplied.12.024037}%
  \BibitemOpen
  \bibfield  {author} {\bibinfo {author} {\bibfnamefont {C.~P.}\ \bibnamefont
  {Blakemore}}, \bibinfo {author} {\bibfnamefont {A.~D.}\ \bibnamefont
  {Rider}}, \bibinfo {author} {\bibfnamefont {S.}~\bibnamefont {Roy}}, \bibinfo
  {author} {\bibfnamefont {A.}~\bibnamefont {Fieguth}}, \bibinfo {author}
  {\bibfnamefont {A.}~\bibnamefont {Kawasaki}}, \bibinfo {author}
  {\bibfnamefont {N.}~\bibnamefont {Priel}}, \ and\ \bibinfo {author}
  {\bibfnamefont {G.}~\bibnamefont {Gratta}},\ }\href {\doibase
  10.1103/PhysRevApplied.12.024037} {\bibfield  {journal} {\bibinfo  {journal}
  {Phys. Rev. Applied}\ }\textbf {\bibinfo {volume} {12}},\ \bibinfo {pages}
  {024037} (\bibinfo {year} {2019}{\natexlab{b}})}\BibitemShut {NoStop}%
\bibitem [{\citenamefont {\relax~microParticles GmbH}()}]{German_sphere}%
  \BibitemOpen
  \bibfield  {author} {\bibinfo {author} {\bibnamefont {\relax~microParticles
  GmbH}},\ }\href@noop {} {}\bibinfo {howpublished}
  {[\url{https://microparticles.de/en}]}\BibitemShut {NoStop}%
\bibitem [{\citenamefont {Moore}\ \emph {et~al.}(2014)\citenamefont {Moore},
  \citenamefont {Rider},\ and\ \citenamefont
  {Gratta}}]{PhysRevLett.113.251801}%
  \BibitemOpen
  \bibfield  {author} {\bibinfo {author} {\bibfnamefont {D.~C.}\ \bibnamefont
  {Moore}}, \bibinfo {author} {\bibfnamefont {A.~D.}\ \bibnamefont {Rider}}, \
  and\ \bibinfo {author} {\bibfnamefont {G.}~\bibnamefont {Gratta}},\ }\href
  {\doibase 10.1103/PhysRevLett.113.251801} {\bibfield  {journal} {\bibinfo
  {journal} {Phys. Rev. Lett.}\ }\textbf {\bibinfo {volume} {113}},\ \bibinfo
  {pages} {251801} (\bibinfo {year} {2014})}\BibitemShut {NoStop}%
\bibitem [{\citenamefont {Abraham}\ and\ \citenamefont
  {Cornell}(1998)}]{ApplOpt.37.1762}%
  \BibitemOpen
  \bibfield  {author} {\bibinfo {author} {\bibfnamefont {E.~R.}\ \bibnamefont
  {Abraham}}\ and\ \bibinfo {author} {\bibfnamefont {E.~A.}\ \bibnamefont
  {Cornell}},\ }\href {\doibase 10.1364/AO.37.001762} {\bibfield  {journal}
  {\bibinfo  {journal} {Appl. Opt.}\ }\textbf {\bibinfo {volume} {37}},\
  \bibinfo {pages} {1762} (\bibinfo {year} {1998})}\BibitemShut {NoStop}%
\bibitem [{\citenamefont {Nieminen}\ \emph {et~al.}(2007)\citenamefont
  {Nieminen}, \citenamefont {Loke}, \citenamefont {Stilgoe}, \citenamefont
  {Knöner}, \citenamefont {Brańczyk}, \citenamefont {Heckenberg},\ and\
  \citenamefont {Rubinsztein-Dunlop}}]{ott}%
  \BibitemOpen
  \bibfield  {author} {\bibinfo {author} {\bibfnamefont {T.~A.}\ \bibnamefont
  {Nieminen}}, \bibinfo {author} {\bibfnamefont {V.~L.~Y.}\ \bibnamefont
  {Loke}}, \bibinfo {author} {\bibfnamefont {A.~B.}\ \bibnamefont {Stilgoe}},
  \bibinfo {author} {\bibfnamefont {G.}~\bibnamefont {Knöner}}, \bibinfo
  {author} {\bibfnamefont {A.~M.}\ \bibnamefont {Brańczyk}}, \bibinfo {author}
  {\bibfnamefont {N.~R.}\ \bibnamefont {Heckenberg}}, \ and\ \bibinfo {author}
  {\bibfnamefont {H.}~\bibnamefont {Rubinsztein-Dunlop}},\ }\href
  {http://stacks.iop.org/1464-4258/9/i=8/a=S12} {\bibfield  {journal} {\bibinfo
   {journal} {J. Opt. A}\ }\textbf {\bibinfo {volume} {9}},\ \bibinfo {pages}
  {S196} (\bibinfo {year} {2007})}\BibitemShut {NoStop}%
\end{thebibliography}%

\end{document}